\documentclass{aastex631}

\graphicspath{{./}{figures/}}
\begin{document}

%\title{Template \aastex Article with Examples: 
%v6.3.1\footnote{Released on March, 1st, 2021}}

\title{Evaluating VLT SPHERE for Probing Diffuse Circumstellar Regions of Evolved Stars}

\author[0009-0006-0547-1030]{Priyadarshini Gokuldass}
% \author{Priyadarshini Gokuldass}
\affiliation{Department of Physical Sciences, College of Arts and Science, Embry-Riddle Aeronautical University, 1 Aerospace Boulevard, Daytona Beach, Fl 32114, USA}

%\collaboration{6}{(AAS Journals Data Editors)}

\author[0000-0002-8348-5191]{Saida M. Caballero-Nieves}
% \author{Saida M. Caballero-Nieves}
\affiliation{Department of Physical Sciences, College of Arts and Science, Embry-Riddle Aeronautical University, 1 Aerospace Boulevard, Daytona Beach, Fl 32114, USA}

\author{Alex Carciofi}
\affiliation{University of S\~ao Paulo, Institute of Astronomy, Geophysics and Atmospherical Sciences 05508-900 S\~ao Paulo, SP, Brazil}

\author[0000-0002-9077-2025]{Bruce Gendre}
% \author{Bruce Gendre}
\affiliation{College of Science and Mathematics, University of the Virgin Islands, 2 John Brewer's Bay, 00802 St Thomas, VI, USA}

\author[0000-0001-5563-6629]{Tajan H. de Amorim}
% \author{Tajan H. de Amorim}
\affiliation{University of S\~ao Paulo, Institute of Astronomy, Geophysics and Atmospherical Sciences 05508-900 S\~ao Paulo, SP, Brazil}

\author[0009-0002-6296-4043]{Keefe J. Kamp}
% \author{Keefe Kamp}
\affiliation{Department of Physical Sciences, College of Arts and Science, Embry-Riddle Aeronautical University, 1 Aerospace Boulevard, Daytona Beach, Fl 32114, USA}

\author[0000-0003-0733-7215]{Justyn Maund}
\affiliation{Physics and Astronomy, Department of Physical Sciences and Engineering, Royal Holloway - University of London, Egham Hill, Egham TW20 0EX UK}

\author[0000-0001-6866-1436]{N. Brice Orange}
% \author{N. Brice Orange}
\affiliation{OrangeWave Innovative Science, LLC, Moncks Corner, SC 29461, USA}
\affiliation{Etelman Observatory Research Center, University of the Virgin Islands, St. Thomas, USVI, USA}

\author[0000-0003-2911-0898]{Maddalena Reggiani}
\affiliation{Institute of Astronomy, KU Leuven, Celestijnenlaan 200D, 3001 Leuven, Belgium}



\begin{abstract}

Red and Yellow Supergiant stars, generally found in late evolutionary stages with high mass loss, provide test cases for evaluating high-contrast imaging performance. Determining how well adaptive optics combined with coronagraphy isolates faint circumstellar emission allows us to define its true sensitivity limits for evolved stars and refine future high-contrast observing strategies. We conducted a pilot study of seven nearby (within 0.8 kpc), bright ($V_{mag} \le 10$) evolved red and yellow supergiants. We selected these targets to evaluate the ability of the VLT's SPHERE/IRDIS high-contrast imaging instrument to detect faint, diffuse circumstellar emission in four infrared filters (J1, J2, K1, and K2). We processed the data using RDI to remove the stellar PSF, AO speckles, and diffraction patterns. We extracted radial profiles and performed Radon transform analysis to search for faint extended emission around each target. We detect no extended circumstellar emission in our high-contrast images or clear infrared excess in the SEDs of our seven targets. Injection-recovery tests demonstrate sensitivity to diffuse emission at contrasts of $\sim \Delta$\,3.5 mag in the J1 and $\sim \Delta$\,2.0 mag in K1 relative to the residual coronagraphic PSF halo. This study provides the first results that elucidate the observational conditions under which SPHERE/IRDIS fails to recover low surface brightness emission and highlights the observational factors that most strongly control detectability, including filter choice, sky calibration, and reference-star matching. Our results motivate future studies to prioritize targets with infrared excess and optimize observing strategies to improve sensitivity to faint circumstellar emission around evolved massive stars.

%These results motivate future studies that combine improved target selection with optimized observing strategies to probe faint circumstellar material around evolved massive stars.

\end{abstract}

% \keywords{\uat{Evolved stars} {481} --- \uat{Circumstellar matter} {241} --- \uat{High contrast techniques} {2369} --- \uat{coronagraphic imaging} {313} --- \uat{Supergiant stars} {1661} --- \uat{Near infrared astronomy} {1093}}

\keywords{
Evolved stars (481) ---
Circumstellar matter (241) ---
High contrast techniques (2369) ---
Coronagraphic imaging (313) ---
Supergiant stars (1661) ---
Near infrared astronomy (1093)
}

\section{High-contrast imaging of circumstellar environments of evolved stars} \label{sec:intro}
Evolved massive stars play an essential role in shaping the universe and enriching the interstellar medium through their winds, eruptions, supernova explosions, and production of heavy elements \citep{Frost_2021}. These processes recycle material into the next generation of stars and planets, profoundly influencing galactic evolution. The geometry, rate, and variability of this mass-loss set up the circumstellar environment into which the star will eventually explode, linking late-stage evolution to the observed properties of its supernova \citep{Matthews2024, Mauron2006}.

Massive stars can shed more than half of their initial mass through their post-main-sequence evolution before they explode as supernova \citep{Stothers1996, VanLoon2008, Beasor2022}; yet, the physical mechanisms that drive this mass-loss-whether steady winds, eruptions, pulsations, convection, magnetic activity, or binary interaction—remain uncertain. These mechanisms are expected to produce distinct geometries and time-variable structures in the circumstellar environment, which high angular resolution observations can directly study \citep{Kervella2016}. Spatially resolving this material on sub-arcsecond scales around bright, nearby evolved stars remains difficult because the emission is intrinsically faint, spatially diffuse, and often dominated by the stellar halo and instrumental PSF. 

Over the past decade, high resolution observations have revealed that the circumstellar environments of cool stars are far from smooth: clumpy dust close to the star \citep{Ohnaka2014}, asymmetric outflows, and localized ejection events \citep{Kervella2016, Cannon2021} have been imaged around red supergiants using a combination of optical/ infrared (IR) interferometry, high-contrast imaging, and polarimetric techniques on 8–10 m class ground-based telescopes equipped with adaptive optics (AO). These studies demonstrate that, when a source is sufficiently dusty, ground-based instruments can resolve the inner tens of milli-arcseconds and reveal newly formed dust structures. By resolving this material, we can directly probe the spatial structure and variability of mass-loss on small spatial scales, providing critical constraints on the physical processes driving stellar evolution.

% Subaru Coronographic Extreme Adaptive Optics 
The advent of extreme adaptive optics (xAO) coronagraphic systems such as Spectro-Polarimetric High-contrast Exoplanet REsearch instrument (SPHERE), Gemini Planet Imager (GPI), Subaru Coronagraphic Extreme Adaptive Optics (SCExAO), and Magellan Adaptive Optics eXtreme (MagAO-X) has made it possible to directly image faint circumstellar material around cool stars \citep{Lomax2025}. Although originally designed for exoplanet studies, these instruments now provide an unprecedented view of the circumstellar environments of massive stars, revealing complex, asymmetric morphologies \citep{Kervella2016, 2015A&A...584L..10S, Cannon2021, Lomax2025}. Similar AO imaging and coronagraphic pilot programs on Gemini and other facilities have resolved arcs, clumps, and dusty knots around very luminous red supergiants such as VY~CMa \citep{2015A&A...584L..10S, Kami2019}, pointing to episodic, non-spherical mass-loss. Among these techniques, polarimetric imaging has become one of the most widely used approaches for probing circumstellar dust around evolved stars.

Polarimetric imaging with instruments such as SPHERE/Zurich Imaging Polarimeter (ZIMPOL) allows constraints on dust grain size, scattering geometries, and asymmetric structures, and remains a central tool in the field. However, these observations are typically time-intensive and optimized for characterizing scattering properties rather than rapidly assessing the presence of faint extended emission. In contrast, our approach focuses on direct coronagraphic imaging, where we aim to detect extended emission and assess instrumental sensitivity rather than characterize scattering properties in detail.

Optical polarimetric imaging with SPHERE/ZIMPOL provides an alternative approach for detecting circumstellar dust through polarized scattered light. However, polarimetric observations require measurements at multiple half-wave-plate positions to obtain the Stokes parameters. As an example, the SPHERE User Manual (Period 97, Phase 2) gives a representative ZIMPOL STOKES QU sequence with a net integration time of 3960 s (66 min), compared with approximately 4.5 min and 16 min for our J- and K-band the Infrared Dual-band Imager and Spectrograph (IRDIS) observations, respectively \citep{sphere_manual_p97}. This corresponds to approximately 15 and 4 times the integration times of our J- and K-band observations. The exact observing time depends on the selected exposure time, number of frames, polarimetric cycles, dithering positions, and field orientations.

The James Webb Space Telescope has extended high-contrast imaging capabilities into the mid-infrared. Its MIRI coronagraph resolved 17 concentric dust shells expanding around the Wolf-Rayet binary WR~140, each tracing distinct mass-ejection episodes over the past century \citep{Lau2022}. Ground-based near-infrared high-contrast imaging and JWST mid-infrared observations probe different components of circumstellar environments. Near-infrared observations with instruments such as SPHERE can probe relatively warm material and scattered light, while JWST/MIRI is more sensitive to thermal emission from cooler dust at longer wavelengths. Combining these wavelength ranges can therefore provide a more complete view of circumstellar material around evolved massive stars. Most of these detections, however, have focused on unusually dusty systems or stars already known to host bright circumstellar shells, and therefore do not test how well these instruments perform on a small, magnitude-limited sample of RSGs and yellow supergiants (YSGs) with no prior evidence for extended emission. Recent high-contrast studies, including GPI detections \citep{Lomax2025}, further demonstrate the ability of high-contrast imaging to probe circumstellar structure around evolved stars.

In this pilot study, we focus on a direct-imaging test case using the SPHERE/IRDIS instrument with classical AO and coronagraphy. This approach is simpler and less telescope-time intensive than full polarimetric campaigns and provides a clean assessment of the instrumental limits for detecting faint, low–surface-brightness circumstellar material. SPHERE/IRDIS combines xAO, a coronagraph, and dual-band near-IR imaging originally developed for exoplanet studies, offering a practical opportunity to evaluate how sensitive this setup is under typical service-mode conditions. Our goal is to determine what observational factors most strongly limit detectability in such data and to assess whether diffuse circumstellar structure can be recovered around bright evolved stars using coronagraph and classical AO imaging.

In order to test SPHERE/IRDIS's ability detecting faint circumstellar emission, we observed a pilot sample of seven nearby RSGs and YSGs (within distance $d \lesssim 0.8$ kpc) in four IRDIS filters (J1, J2, K1, K2). The targets were selected for their brightness and proximity, ensuring that SPHERE’s spatial resolution could in principle resolve structures on scales of a few hundred AU (Section~\ref{sec:sample}). We processed the observations using Reference Differential Imaging (RDI), followed by radial-profile measurements and Radon transform analysis against theoretical point-spread functions (PSFs) to search for faint extended emission (Section~\ref{sec:analysis}). In parallel, we built spectral energy distributions (SEDs) for all stars using archival data to check for any IR excess indicative of circumstellar dust. The resulting non-detections are discussed in the context of SPHERE’s current sensitivity limits and observational constraints relevant for future SPHERE studies (Section~\ref{sec:discussion}).

\section{Sample Selection and SPHERE/IRDIS Observations} \label{sec:sample}
 
We selected seven nearby red and yellow Supergiants that are bright enough ($V \lesssim 10$ mag) for SPHERE’s Natural Guide Star mode and close enough ($d \lesssim 0.8$ kpc) that any circumstellar material would be spatially resolvable above the diffraction limit. Our observations used four IR filters (J1, J2, K1, and K2), sampling the near-IR regime where scattered light and circumstellar emission dominate, to test SPHERE/IRDIS sensitivity to low–surface-brightness structures.

While other AO instruments—such as VLTI/PIONIER, VLT/NACO, or Gemini/NIRI—can achieve either high angular resolution or high contrast, SPHERE uniquely combines both, by employing extreme AO with a ALC\_YJH\_S apodized Lyot coronagraph, making it an ideal instrument for exploring faint, diffuse circumstellar emission around luminous evolved massive stars. We obtained our data using the IRDIS instrument \citep{2008SPIE.7014E..3LD, 2010MNRAS.407...71V, 2019A&A...631A.155B} which is designed primarily for high-contrast exoplanet detection through pupil tracking and differential imaging techniques. In this mode, the field rotates with the parallactic angle while the telescope pupil remains fixed, enabling speckle suppression through differential imaging process (see section 3.2)(\citealt{2008SPIE.7014E..3LD, de2020polarimetric}). 

Our sample includes five Yellow Supergiants (YSGs) and two RSGs and are listed in Table \ref{Sample}. The angular radii listed in Table~\ref{Sample} were calculated from the adopted stellar radii and Bailer-Jones geometric distances \citep{BailerJones2021} and are not direct interferometric measurements. These stars are among the brightest massive evolved stars accessible for adaptive optics observations, all brighter than the R $\leq$ 11 mag limit required for AO guiding. Their spectral types span A5–M1, with $V$ magnitudes ranging from 2.9 to 10.0 mag.

We observed the seven targets for a total of 8.6 hours using pupil tracking mode with SPHERE/IRDIS in the J and K bands. These bands are subdivided into J1 (1.19 $\mu$m), J2 (1.27 $\mu$m), K1 (2.10 $\mu$m), and K2 (2.24 $\mu$m), based on their central wavelengths. Coronagraphic exposures provided the science frames, while short non-coronagraphic exposures supplied unsaturated PSFs for calibration. The coronagraph effectively suppressed the stellar PSF, particularly within $\sim$1$''$, improving contrast and isolating potential extended emission. The observing conditions for all targets are summarized in Table \ref{Obs-detail}.

The instrument specifications and filter parameters are summarized in Table \ref{Specs}. We calculated the expected AO correction radius, which defines the approximate angular extent of the region corrected by the adaptive optics system, as \(20\lambda/D\) for each filter. We adopted the coronagraph inner working angle (IWA) from the SPHERE User Manual (Period 97, Phase 2) \cite{sphere_manual_p97}. 

\begin{figure}[h]
    \centering
    \includegraphics[width=\textwidth]{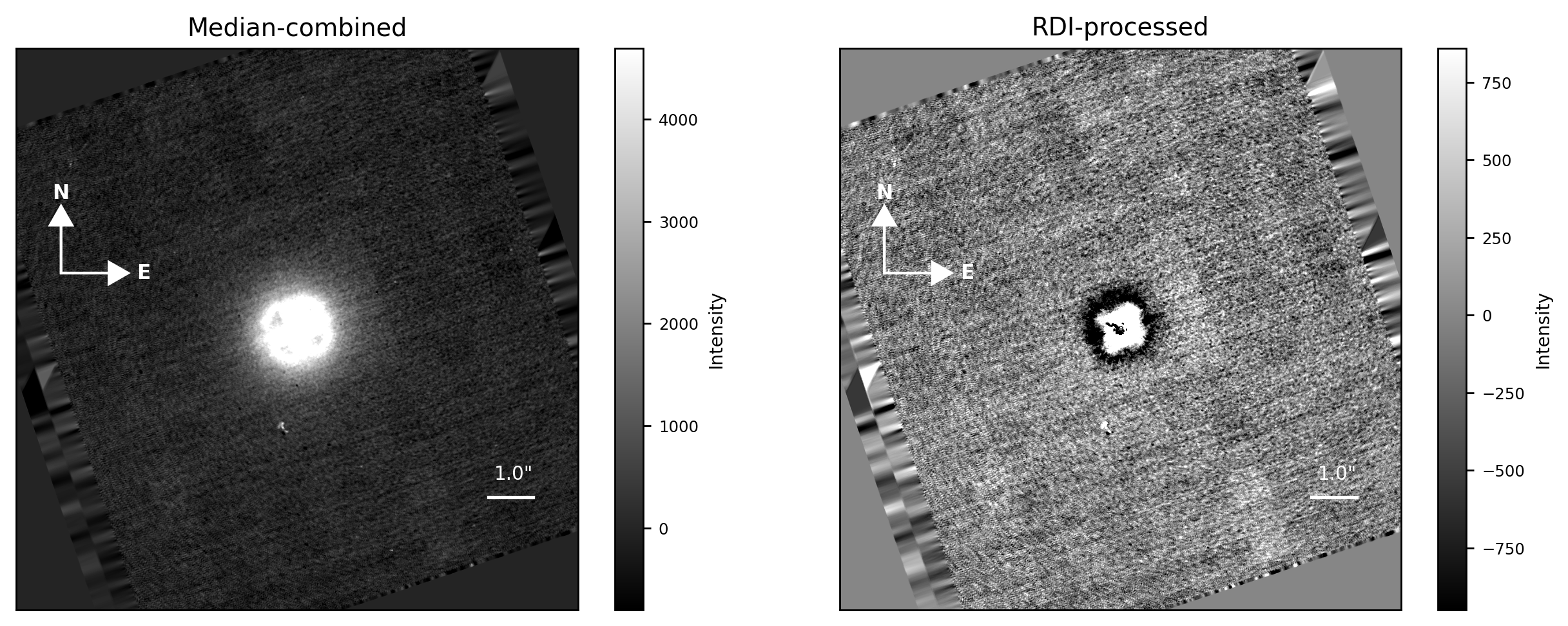}
    \caption{Example of comparison between the median-combined (left) and RDI-processed (right) images (here HD~74395) in the \textit{J1} filter. The median-combined frame represents the coronagraphic stellar PSF and diffraction pattern prior to reference differential imaging, while the RDI-processed image shows the residuals after reference PSF subtraction. We do not detect any localized or asymmetric intensity enhancements that would indicate the presence of extended circumstellar emission beyond the expected coronagraphic halo and diffraction pattern. North and East are indicated in the image, and the white bar corresponds to 1~arcsec.
    \label{fig:HD74395_J1_RDI}}
\end{figure}

\begin{deluxetable}{lcclccccc}
\tablenum{1}
\tablecaption{
RSG and YSG sample observed in this pilot study. Columns list target name, right ascension ($\alpha$), declination ($\delta$), spectral type, $V$ magnitude, geometric distance from \citet{BailerJones2021} ($d_{\rm BJ21}$), effective temperature ($T_{\rm eff}$), stellar radius ($R_\star$), and angular stellar radius ($\theta$). \label{Sample}}
\tablewidth{0pt}

\tablehead{
\colhead{Target} &
\colhead{$\alpha$ [J2000] (h)} &
\colhead{$\delta$ [J2000] (deg)} &
\colhead{Spectral Type} &
\colhead{$V$ mag.} &
\colhead{$d_{\rm BJ21}$ (kpc)} &
\colhead{$T_{\mathrm{eff}}$ (K)} &
\colhead{$R_\star/R_\odot$} &
\colhead{$\theta$ (mas)}
}

\startdata
HD 38915  & 05:49:00 & $-03$:18:41 & A5\,Ib   & 9.87 & $0.328^{+0.003}_{-0.003}$ & 8100 & \nodata & \nodata \\
HD 50877  & 06:54:08 & $-24$:11:03 & K2.5\,I  & 3.87 & $0.783^{+0.090}_{-0.073}$ & 3900 & 280   & 1.66 \\
HD 52877  & 07:01:43 & $-27$:56:05 & M1.5\,I  & 3.47 & $0.405^{+0.030}_{-0.028}$ & 3750 & 420   & 4.82 \\
HD 58526  & 07:25:51 & $-05$:46:30 & G3\,Ib   & 5.96 & $0.572^{+0.008}_{-0.009}$ & 5380 & 28.19 & 0.23 \\
HD 67594  & 08:08:36 & $-02$:59:02 & G2\,Ib   & 4.33 & $0.309^{+0.013}_{-0.013}$ & 5210 & 56.73 & 0.85 \\
HD 74395  & 08:43:40 & $-07$:14:01 & G1\,Ib   & 4.62 & $0.296^{+0.031}_{-0.026}$ & 5370 & 34.43 & 0.54 \\
HD 209750 & 22:05:47 & $-00$:19:11 & G2\,Ib   & 2.94 & $0.202^{+0.020}_{-0.016}$ & 5210 & 55.65 & 1.28 \\
\enddata

\tablecomments{We obtained the coordinates and $V$ magnitudes from SIMBAD. We obtained the geometric distance estimates and their lower and upper uncertainty bounds from \citet{BailerJones2021}, based on Gaia EDR3 parallaxes. We adopted the spectral type and effective temperature of HD 38915 from \citet{Pecaut2013}. We adopted the effective temperatures and stellar radii of HD 50877 and HD 52877 from \citet{Levesque2005}. For HD 58526, HD 67594, HD 74395, and HD 209750, we adopted the effective temperatures from \citet{Lyubimkov2010}, and calculated the stellar radii from the masses and surface gravities reported by \citet{Lyubimkov2010}, using $R_\star=\sqrt{GM_\star/g}$. We calculated the angular stellar radii from the adopted stellar radii and the geometric distances using $\theta=R_\star/d$.
}

\end{deluxetable}

\begin{deluxetable}{lcccc|cccc}
\tablenum{2}
\tablecaption{Log of observations, including observing dates (yyyy/mm/dd), total exposure time ($t_{\mathrm{exp}}$) in minutes and conditions in J and K band for the sample targets. All targets were observed for 0.83s per image in the cube. A ND\_2 neutral density filter was used for all targets except HD 38915 marked by $^{*}$, which was observed with ND\_1 filter. \label{Obs-detail}}
\tablewidth{0pt}
\tablehead{
\colhead{Target} &
\multicolumn{4}{c}{J Band} &
\multicolumn{4}{c}{K Band} \\
\colhead{} &
\colhead{Obs. Date} &
\colhead{Airmass ($z$)} &
\colhead{Seeing ($''$)} &
\colhead{$t_{\mathrm{exp}}$ (min)} &
\colhead{Obs. Date} &
\colhead{Airmass ($z$)} &
\colhead{Seeing ($''$)} &
\colhead{$t_{\mathrm{exp}}$ (min)}
}

% \decimalcolnumbers
\startdata
HD 38915$^{*}$  & 2017/03/15 & $1.373-1.431$ & $0.54-0.67$ & 4.47 & 2017/02/16 & $1.287-1.465$ & $0.43-0.53$ & 8.93\\ 
HD 50877        & 2016/10/21 & $1.035-1.400$ & $1.51-2.05$ & 4.47 & 2017/03/05 & $1.067-1.410$ & $0.91-1.47$& 16.08\\ 
HD 52877        & 2017/01/09 & $1.020-1.025$ & $0.36-1.82$ & 4.47 & 2017/01/09 & $1.003-1.098$ & $0.39-1.27$& 3.27\\ 
HD 58526        & 2016/10/21 & $1.057-1.058$ & $1.06-1.13$ & 2.23 & 2017/03/07 & $1.058-1.079$ & $0.44-0.66$& 16.08\\ 
HD 67594        & 2016/10/21 & $1.528-1.612$ & $0.81-0.92$ & 4.47 & 2017/02/16 & $1.092-1.139$ & $0.39-0.83$& 16.08\\ 
HD 74395        & 2016/10/21 & $1.567-1.633$ & $0.95-1.23$ & 4.47 & 2017/02/15 & $1.048-1.063$ & $0.71-1.44$ & 16.08\\ 
HD 209750       & 2016/10/05 & $1.172-1.195$ & $0.69-1.26$ & 4.47 & 2016/10/06 & $1.112-1.158$ & $0.69$ & 16.08 \\ 
\enddata
\end{deluxetable}

\begin{deluxetable}{cccccc}
\tablenum{3}
\tablecaption{ Properties of the IRDIS filters used in our SPHERE observations. We list the filter name, wavelength bandpass ($\mu$m), central wavelength $\lambda_c$ ($\mu$m), the AO correction radius in detector pixels, the inner working angle (IWA) in arcseconds, and the coronagraph mask radius in pixels. \label{Specs}}
\tablewidth{0pt}
\tablehead{
\colhead{ Filters } & \colhead{Bandpass} & \colhead{$\lambda_c$} & \colhead{AO correction} &
\colhead{IWA} & \colhead{Coronagraph}\\
\nocolhead{ Filters } & \colhead{($\mu$m)} & \colhead{($\mu$m)} & \colhead{(pixels)} &
\colhead{($''$)} & \colhead{(pixels)}
}
\startdata
 J1 & 1.125 - 1.365 & 1.19 & 50.090 & 0.15 & 12.245  \\
 J2 & 1.125 - 1.365 & 1.27 & 53.460 & 0.15 & 12.245  \\
 K1 & 2.032 - 2.332 & 2.10 & 88.399  & 0.20  & 16.32 \\
 K2 & 2.032 - 2.332 & 2.24 & 94.460 & 0.20  & 16.32 \\
\enddata
\end{deluxetable}

\section{Experimental Design: Testing SPHERE's Capability for Detecting Diffuse Circumstellar Material} \label{sec:analysis}

While the observations of this sample were completed during a single night, the data processing and reduction presented several challenges. The SPHERE High-Contrast Data Centre (HC-DC) \citep{Delorme2017, Vigan2017} performed the initial data reduction, including flat-field correction, bad-pixel removal, and normalization for the neutral density (ND) filters used during the observations. After obtaining the reduced data cubes, we applied a series of differential imaging techniques to search for circumstellar emission, as detailed below.

\subsection{Data Reduction and Differential Imaging}

To identify residual structures and assess their origin, we adopted differential imaging using the Vortex Image Processing (VIP) package \citep{Gomez_Gonzalez_2017, christiaens2023vip, Marois_2006}. VIP provides modular routines for PSF extraction, frame registration, derotation, median combination, and differential imaging analysis, along with tools to calculate signal-to-noise ratio (SNR) maps. The package accepts pre-calibrated 3D FITS data cubes and their associated calibration files (wavelength, parallactic angles, and PSF images) and measures the full width at half maximum by fitting a 2D Gaussian to the core of the unsaturated PSF. 

We obtained the observations in pupil-tracking mode, which enables speckle subtraction through differential imaging techniques. We initially processed the data using Angular Differential Imaging (ADI), but did not identify any obvious circumstellar emission. Because ADI can partially self-subtract extended or symmetric circumstellar structures as the field rotates, we subsequently explored RDI, following a similar approach to \citet{2022A&A...666A..32X}. RDI uses an independent reference star to model and subtract the stellar PSF, allowing better preservation of symmetric circumstellar structures that may be affected by self-subtraction in ADI. We therefore used RDI for the subsequent analysis to search for faint, spatially extended circumstellar emission. We did not identify any clear circumstellar emission with RDI.

Once we processed each calibrated cube using the VIP RDI procedure, we then binned the data into ten combined frames per target to balance computational efficiency and signal stability. We compared the processed frames directly with the median-combined calibrated image to identify any residual features that might represent circumstellar material (see Fig. \ref{fig:HD74395_J1_RDI}). Since these images are background and PSF subtracted, the resulting radial profiles represent relative residual intensities rather than absolute flux measurements. Small negative baseline offsets visible in the radial profile (see Fig. \ref{HD74395-composite-Radial}) can therefore arise from modest over subtraction or background mismatch, particularly in K band where thermal background variations are larger. These offsets are a data-processing artifact and do not correspond to negative astrophysical flux.

Fig. \ref{fig:HD74395_J1_RDI} compares the median-combined and RDI-processed frames of an example star (here HD 74395) in the J1 band. The median image mainly shows the bright coronagraphic halo and diffraction pattern, while the RDI subtraction effectively removes the stellar core and suppresses static speckles. We detect no obvious diffuse emission or asymmetric structure above the noise level. The residual features are consistent with instrumental and detector artifacts rather than astrophysical circumstellar material. Within these limitations, we do not detect statistically significant diffuse circumstellar material in any of our targets.

Although we did not detect any diffuse circumstellar emission, the sensitivity of the observations is limited by several instrumental and observational factors. These include the use of ND filters, short exposure times, the lack of dedicated K-band sky frames, and imperfect reference-star matching for RDI. The impact of these limitations on the detectability of faint extended emission is discussed in Section~\ref{sec:discussion}.

\subsection{Radial profile and Model-Based Radon analysis of HD~74395}
\label{Radon-sinograms}

\begin{figure}[h]
    \centering

    %------------------ J1 ------------------%
    \includegraphics[width=\textwidth]{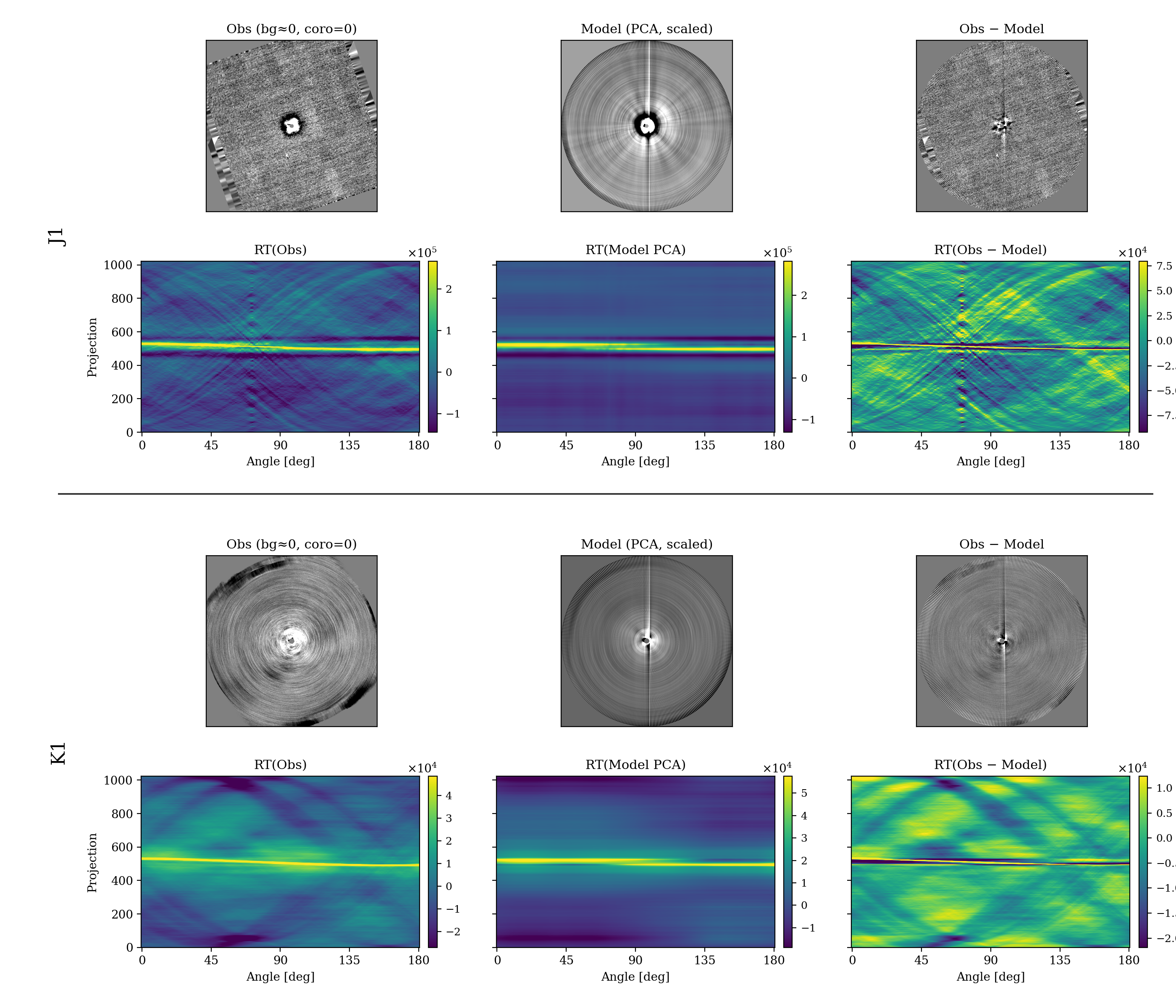}

    \caption{Composite PCA-Radon comparison for HD~74395 in the J1 (top) and K1 (bottom) filters. In each panel, the top row shows the observed coronagraphic image (left), the model reconstructed from the PCA decomposition in Radon space (middle), and the residual image obtained by inverse Radon transform of the residual sinogram (right). The bottom row shows the Radon sinogram of the observed image, the PCA model sinogram, and the residual sinogram obtained by subtracting the model from the observed data in projection space. In both filters, the observed sinograms display the dominant horizontal band associated with the symmetric stellar halo together with faint higher-order diffraction structures. The PCA models reproduce the main symmetric halo structure, while the residuals contain only low-level speckle and diffraction artifacts. Neither the image nor the Radon sinograms show asymmetric or localized structures expected from real circumstellar emission.}

    \label{HD74395-composite-RT}
\end{figure}

Since the initial RDI processing did not reveal clear diffuse emission signatures, we explored several modeling approaches to better characterize the coronagraphic halo structure and assess whether weak diffuse components could be isolated in the data. To separate the dominant axisymmetric halo from potential asymmetric features, we applied a projection-based analysis using the Radon transform \citep{Barrett1982O, Nir2018} to all targets, obtaining consistent results across the sample. In this section, we use HD~74395 as a representative example to illustrate the method. As an initial step, we examined the RDI-processed images and their radial profiles to identify deviations from the expected coronagraphic PSF structure and evaluate whether any extended circumstellar features are present. We then analyzed the corresponding Radon sinogram, where the central bright horizontal band represents the dominant circular halo structure, while fainter sinusoidal features trace higher-order diffraction patterns and noise (see Figures~\ref{HD74395-composite-RT}).

\begin{figure}[h]
    \centering

    %------------------ J1 ------------------%
    \includegraphics[width=\textwidth]{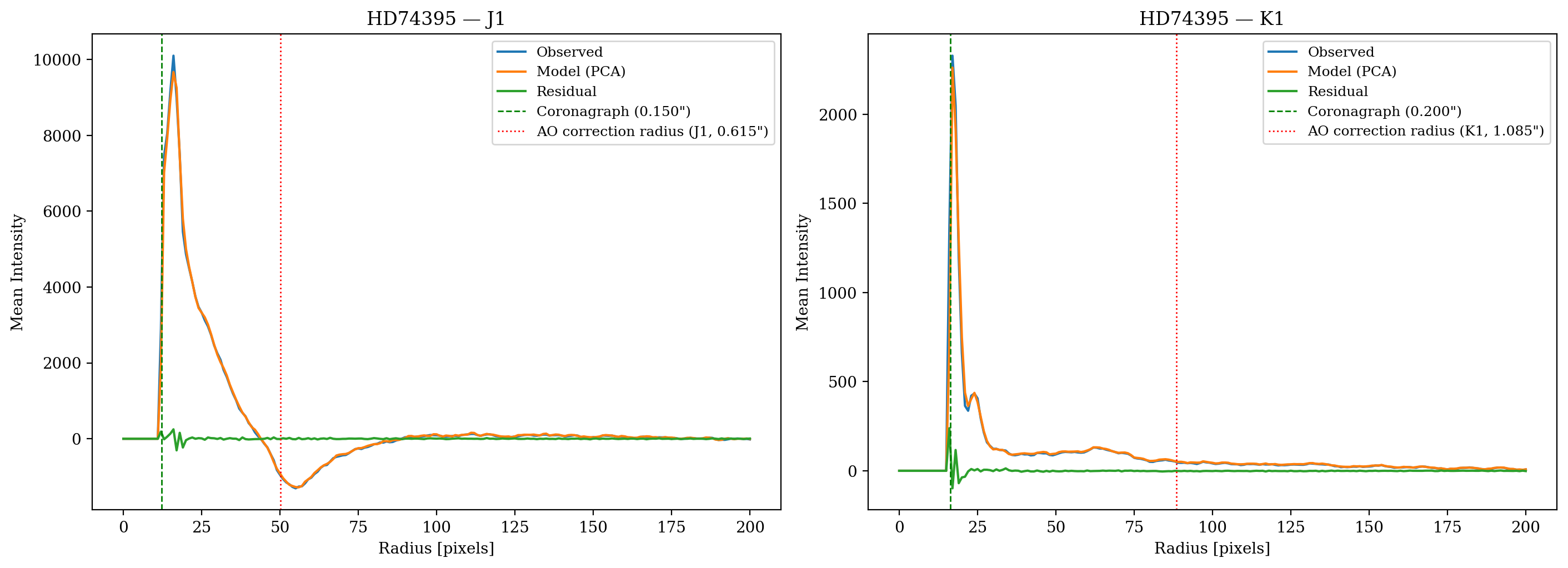}

    \caption{Azimuthally averaged radial intensity profiles for HD~74395 in the J1 (left) and K1 (right) filters after background subtraction and coronagraph masking. In both panels, the blue curve shows the observed coronagraphic image, the orange curve shows the PCA-based Radon-space model reconstructed using the dominant projection modes, and the green curve shows the residual profile obtained from the difference between the observed and modeled data. The dashed vertical lines mark the coronagraph radii (0.150$''$ for J1 and 0.200$''$ for K1), while the dotted vertical lines indicate the AO correction radii} (0.615$''$ for J1 and 1.085$''$ for K1). The close agreement between the observed and modeled profiles demonstrates that the PCA model reproduces the symmetric halo structure well in both filters. Outside the first Airy ring, the residual profiles remain flat and noise-like, with no systematic positive excess or extended structure that would indicate diffuse circumstellar emission in either band.

    \label{HD74395-composite-Radial}
\end{figure}

We then tested several analytical halo models, including Airy functions, Moffat+Gaussian combinations, and circularly symmetric profiles \citep{Bendinelli1988, Trujillo2001} to reproduce the instrumental effects on the images. These models were limited by their functional forms and did not reproduce the observed coronagraphic halo consistently across all filters. We therefore adopted a Radon transform based workflow that represents the image in projection space and isolates the dominant symmetric halo component. In this framework, any additional structure such as diffuse circumstellar emission would appear as localized deviations from the primary horizontal band, enabling a sensitive assessment of low-level residual structure \text{(see Fig. \ref{HD74395-composite-Radial})}.

We finally constructed a principal-component analysis \citep{Yuan1999, Valleti2020} based Radon-space model of the axisymmetric halo by decomposing the observed sinogram and retaining only its dominant modes. This approach captures the coherent circular halo structure without assuming any analytic form. We scaled the reconstructed low-rank sinogram to match the central projection band and generated a model for each filter using the inverse Radon transform. We subtracted the PCA model sinogram from the observed sinogram in projection space and applied the inverse Radon transform to obtain the residual image, in which any extended emission would be clearly visible.

Among all approaches explored, the PCA Radon-space model consistently performs best, accurately reproducing the instrumental response of SPHERE/IRDIS and yielding clean, symmetric residuals. As shown in Fig.~\ref{HD74395-composite-RT}, the model captures the Radon transform features associated with the instrumental halo. The azimuthally averaged radial intensity profiles (Fig.~\ref{HD74395-composite-Radial}) further support this result, with the observed and PCA-modeled profiles exhibiting consistent radial behavior. The online figure sets associated with Figures~\ref{HD74395-composite-RT} and \ref{HD74395-composite-Radial} present the corresponding results for the full sample. Neither the image-domain residuals nor the Radon sinograms show asymmetric or off-axis structures that would indicate detectable extended circumstellar emission.

\begin{figure}[h]
    \centering
    \includegraphics[width=\textwidth]{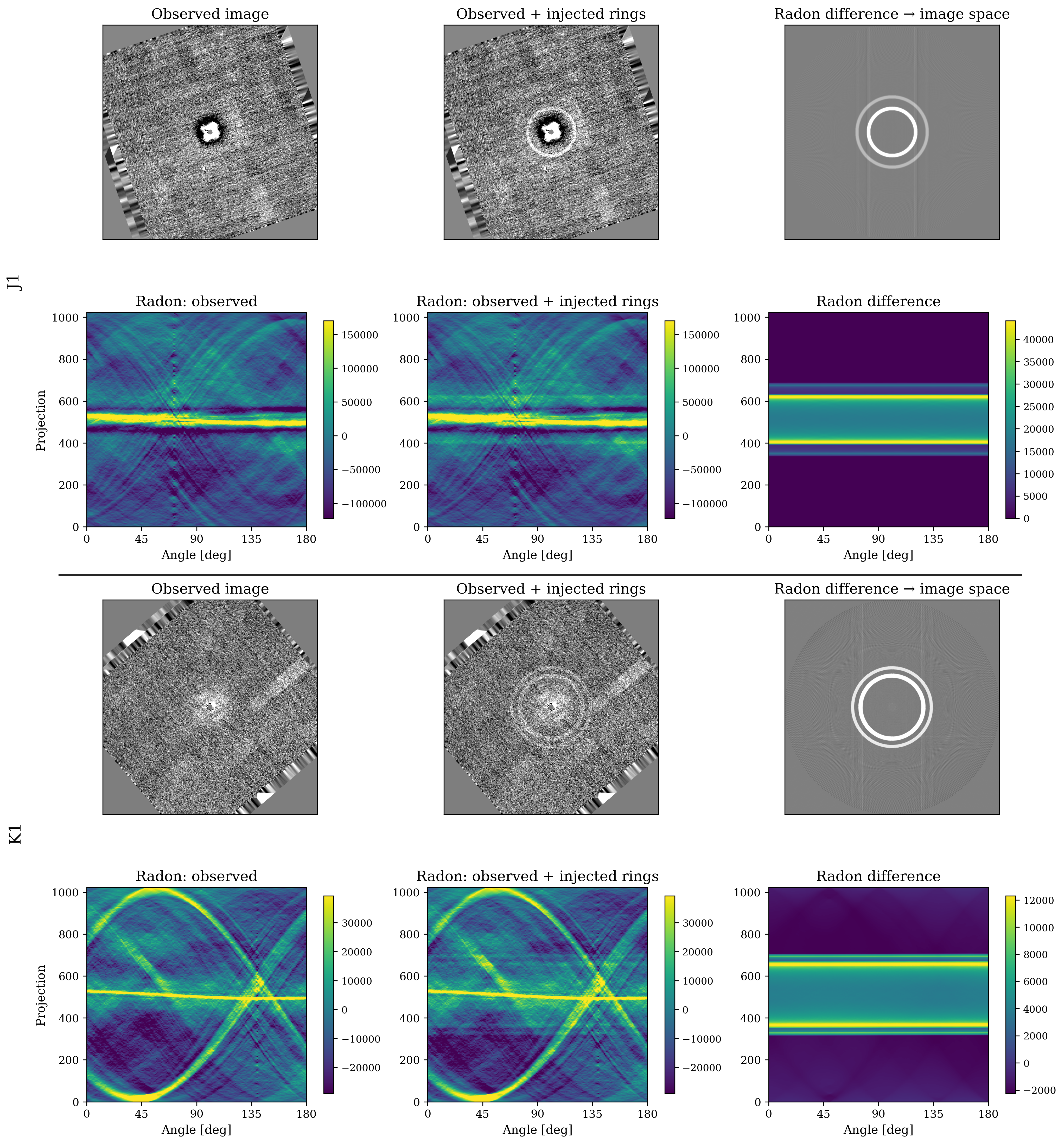}
    \caption{ Injection-recovery tests for HD~74395 in the J1 (top) and K1 (bottom) bands. For each filter, the upper panels show the RDI-processed image without injection (left), the image with two injected rings (middle), and the Radon-space difference mapped back into image space (right); the panels directly below show the corresponding Radon sinograms. In J1, the inner and outer rings correspond to contrasts of 2.0 and 3.5 magnitudes relative to the local residual coronagraphic halo, respectively. In K1, the corresponding injected contrasts are 1.8 and 2.0 magnitudes. The injected structures produce clear, localized signatures in Radon space despite being difficult to distinguish directly in the RDI-processed images. These tests indicate approximate sensitivity limits of $\Delta m \sim 3.5$ mag in J1 and $\Delta m \sim 2.0$ mag in K1 relative to the residual coronagraphic halo.}
    \label{HD74395-J1-fake-ring}
\end{figure}

To quantify the sensitivity of our observations to faint, diffuse circumstellar emission, we performed injection–recovery tests on the J1- and K1-band data for HD~74395 (see Fig. \ref{HD74395-J1-fake-ring}). We injected synthetic ring-like structures into the science data at increasing contrast levels relative to the residual coronagraphic PSF halo and processed the injected and uninjected data using identical RDI and Radon transform procedures.

For the this test, we selected 20 frames evenly spaced throughout the science cube to sample the full observation while reducing the computational cost of the injection-recovery analysis. We used this same 20-frame subset and identical RDI parameters for the injected and uninjected cases. In contrast, we used the complete science cube to produce the K1 RDI image shown in Fig.~\ref{HD74395-composite-RT}. Consequently, the K1 RDI image in Fig.~\ref{HD74395-J1-fake-ring} differs in appearance from the K1 RDI image in Fig.~\ref{HD74395-composite-RT} because the two reductions use different numbers of science frames.

We evaluated detectability both in image space and in Radon space, where circular structures produce distinct signatures in the sinogram. In J1, we recover injected structures down to a contrast of $\sim\Delta$\,3.5 mag relative to the residual coronagraphic PSF halo. In K1, we recover injected structures down to $\sim\Delta$\,2.0 mag. In both bands, the injected structures produce clear signatures in Radon space even when they are difficult to distinguish directly in the RDI-processed images. While this limit is derived for a representative target, we expect similar sensitivity across the sample given the comparable observing conditions and data quality.

Outside the AO correction radius, the residual profiles remain approximately flat and noise-like. At smaller separations, residual stellar-halo structure and subtraction artifacts limit sensitivity to very low-surface-brightness material. The implications of these observational limitations for the detectability of diffuse circumstellar emission are discussed further in Section \ref{sec:discussion}

The online figure set associated with Figure~\ref{HD74395-composite-RT} presents the corresponding PCA-Radon results for all targets. All targets show consistent behavior, and we detect no significant extended emission in any of the sample stars. These results indicate that HD~74395 is representative of the full sample.

% ============================================================
% ONLINE FIGURE SET 2
% ============================================================
\figsetstart
\figsetnum{2}
\figsettitle{Composite PCA-Radon comparison}

\figsetgrpstart
\figsetgrpnum{2.1}
\figsetgrptitle{HD 74395}
\figsetplot{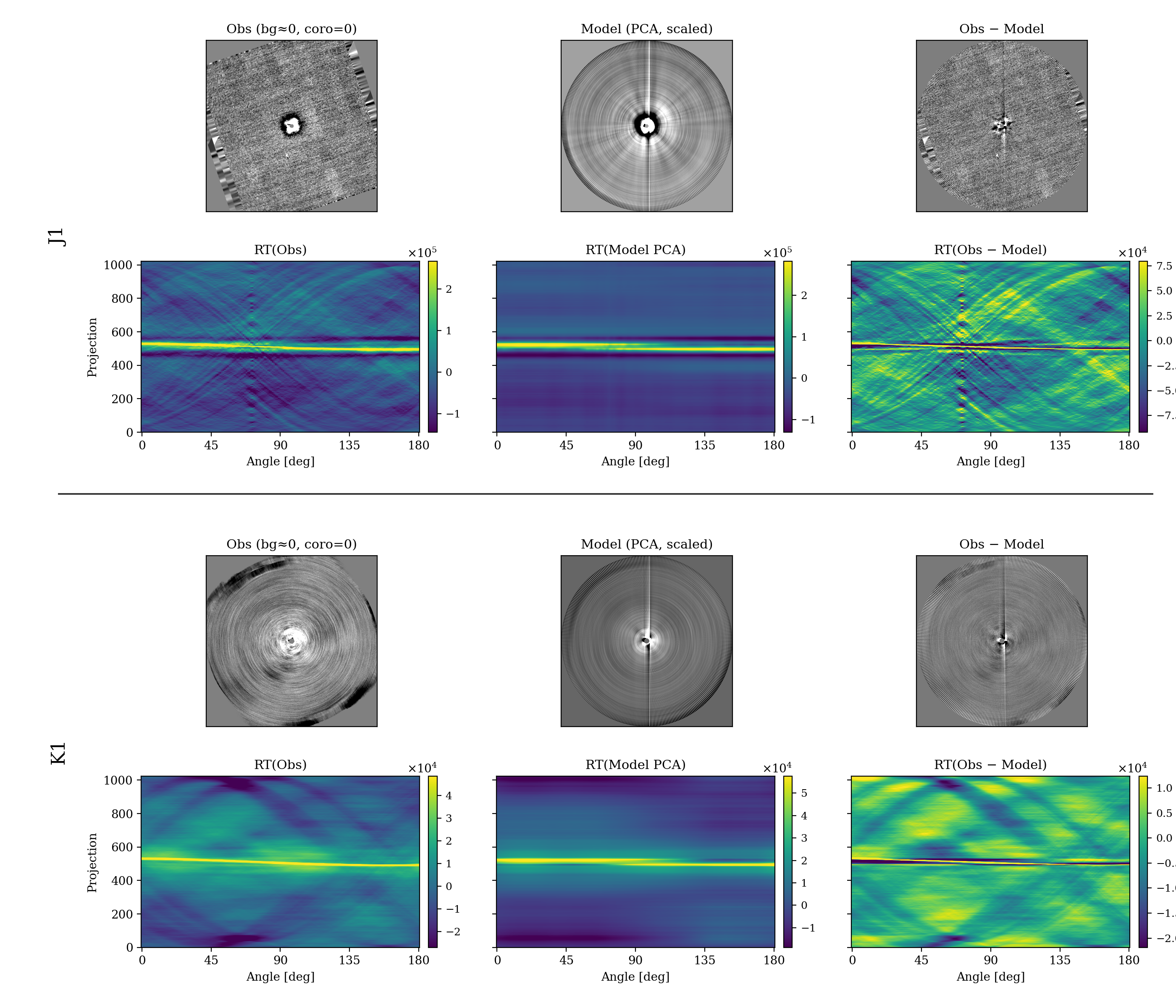}
\figsetgrpnote{Composite PCA-Radon comparison in the J1 (top) and K1 (bottom) filters. In each panel, the top row shows the observed coronagraphic image (left), the model reconstructed from the PCA decomposition in Radon space (middle), and the residual image obtained by inverse Radon transform of the residual sinogram (right). The bottom row shows the Radon sinogram of the observed image, the PCA model sinogram, and the residual sinogram obtained by subtracting the model from the observed data in projection space.}
\figsetgrpend

\figsetgrpstart
\figsetgrpnum{2.2}
\figsetgrptitle{HD 38915}
\figsetplot{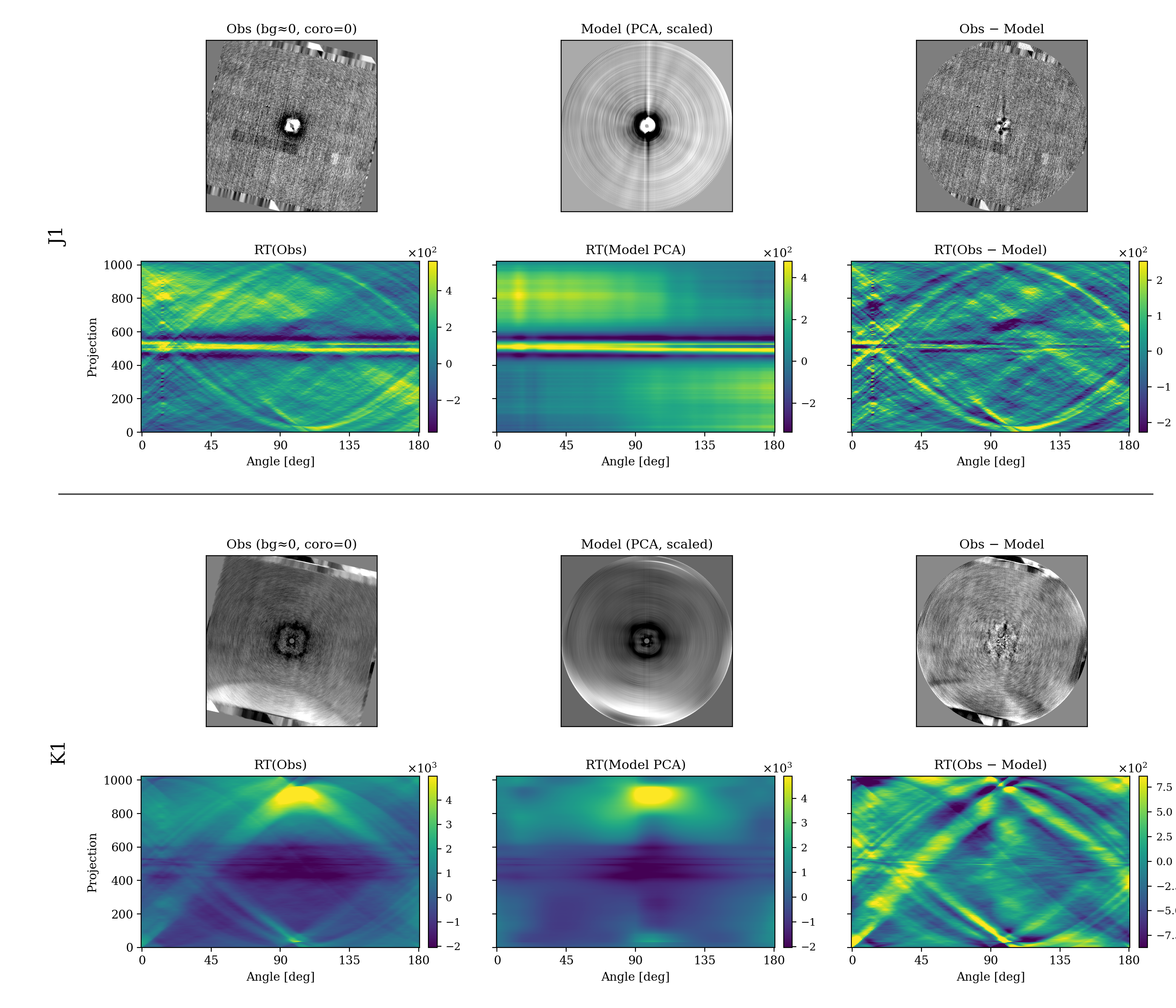}
\figsetgrpnote{Composite PCA-Radon comparison in the J1 (top) and K1 (bottom) filters. In each panel, the top row shows the observed coronagraphic image (left), the model reconstructed from the PCA decomposition in Radon space (middle), and the residual image obtained by inverse Radon transform of the residual sinogram (right). The bottom row shows the Radon sinogram of the observed image, the PCA model sinogram, and the residual sinogram obtained by subtracting the model from the observed data in projection space.}
\figsetgrpend

\figsetgrpstart
\figsetgrpnum{2.3}
\figsetgrptitle{HD 50877}
\figsetplot{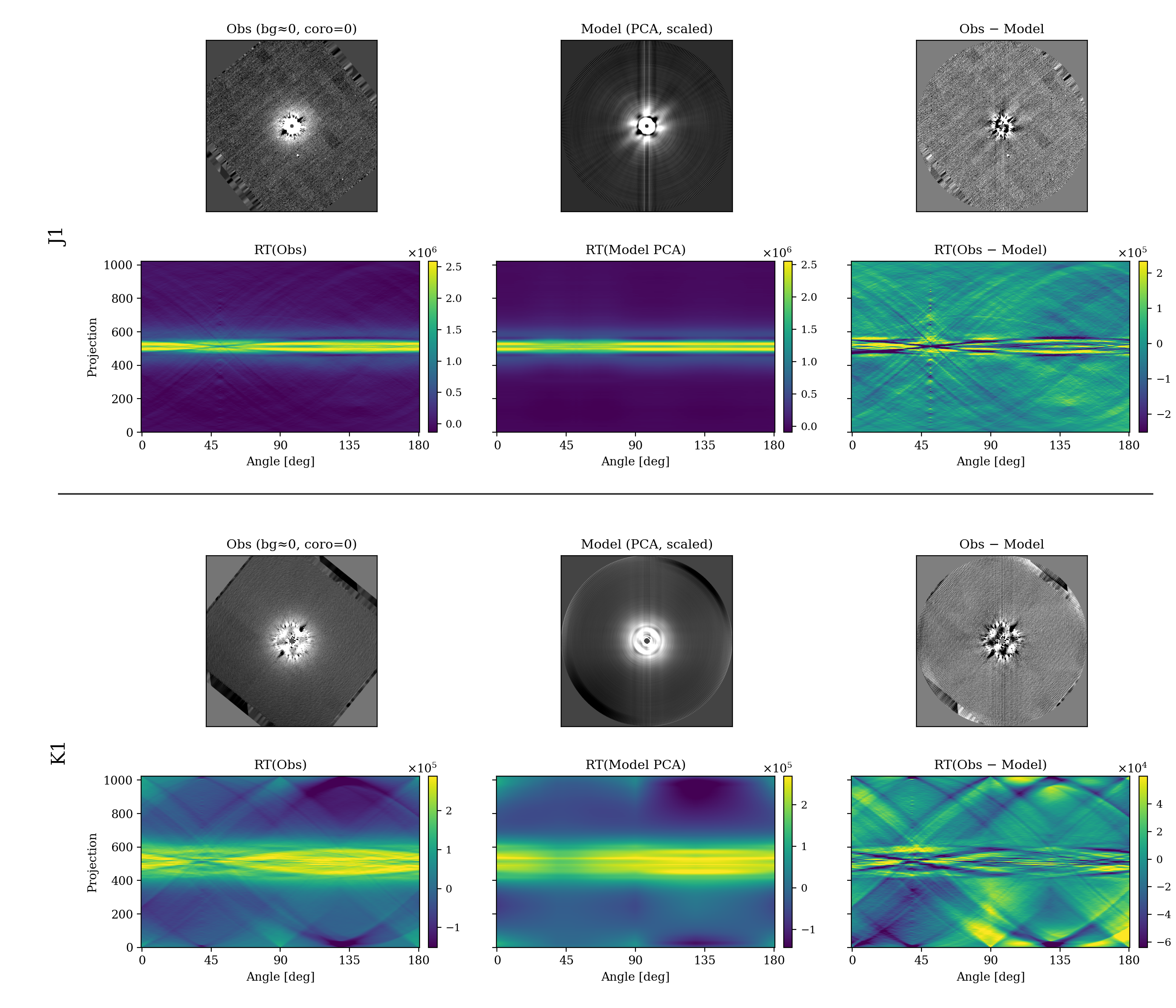}
\figsetgrpnote{Composite PCA-Radon comparison in the J1 (top) and K1 (bottom) filters. In each panel, the top row shows the observed coronagraphic image (left), the model reconstructed from the PCA decomposition in Radon space (middle), and the residual image obtained by inverse Radon transform of the residual sinogram (right). The bottom row shows the Radon sinogram of the observed image, the PCA model sinogram, and the residual sinogram obtained by subtracting the model from the observed data in projection space.}
\figsetgrpend

\figsetgrpstart
\figsetgrpnum{2.4}
\figsetgrptitle{HD 52877}
\figsetplot{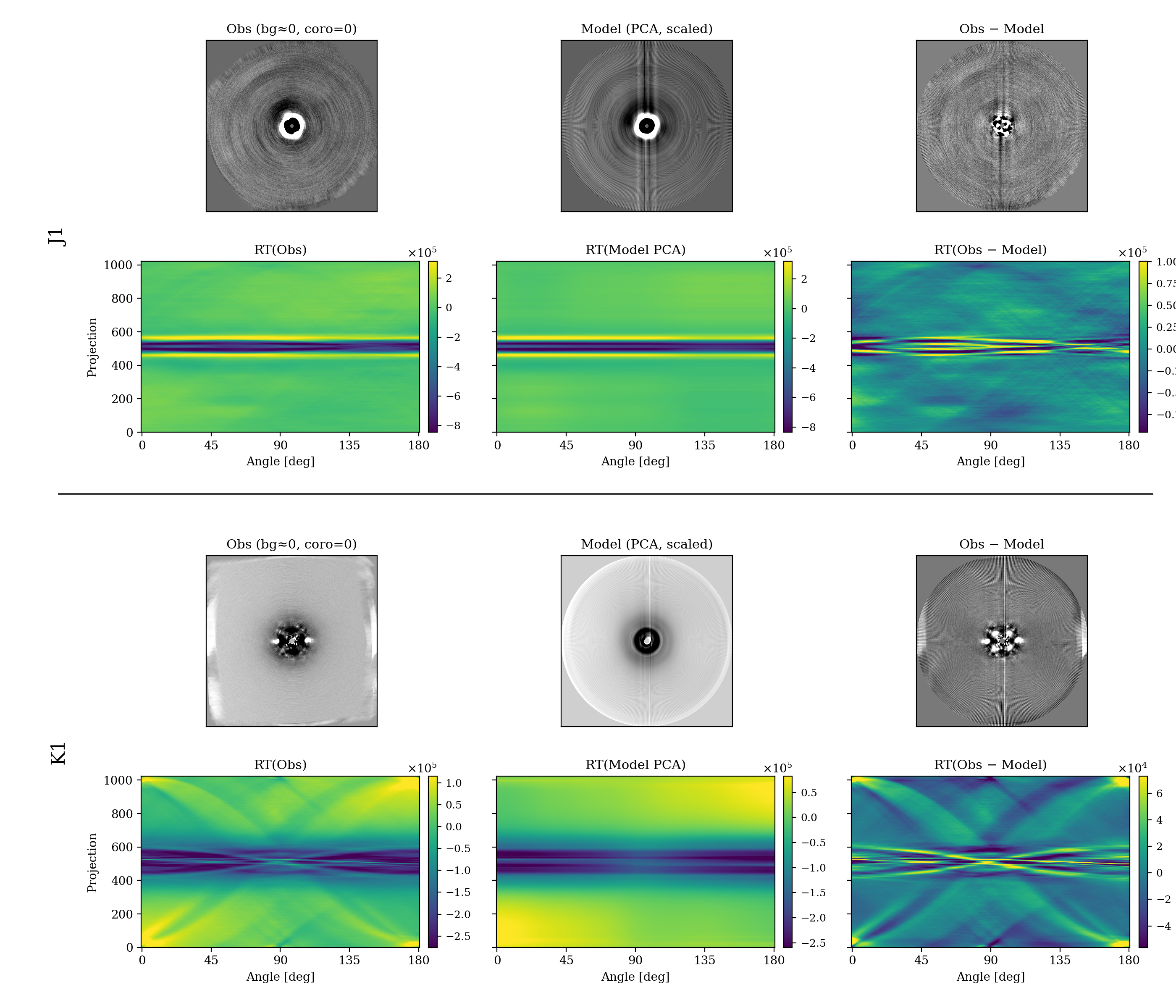}
\figsetgrpnote{Composite PCA-Radon comparison in the J1 (top) and K1 (bottom) filters. In each panel, the top row shows the observed coronagraphic image (left), the model reconstructed from the PCA decomposition in Radon space (middle), and the residual image obtained by inverse Radon transform of the residual sinogram (right). The bottom row shows the Radon sinogram of the observed image, the PCA model sinogram, and the residual sinogram obtained by subtracting the model from the observed data in projection space.}
\figsetgrpend

\figsetgrpstart
\figsetgrpnum{2.5}
\figsetgrptitle{HD 58526}
\figsetplot{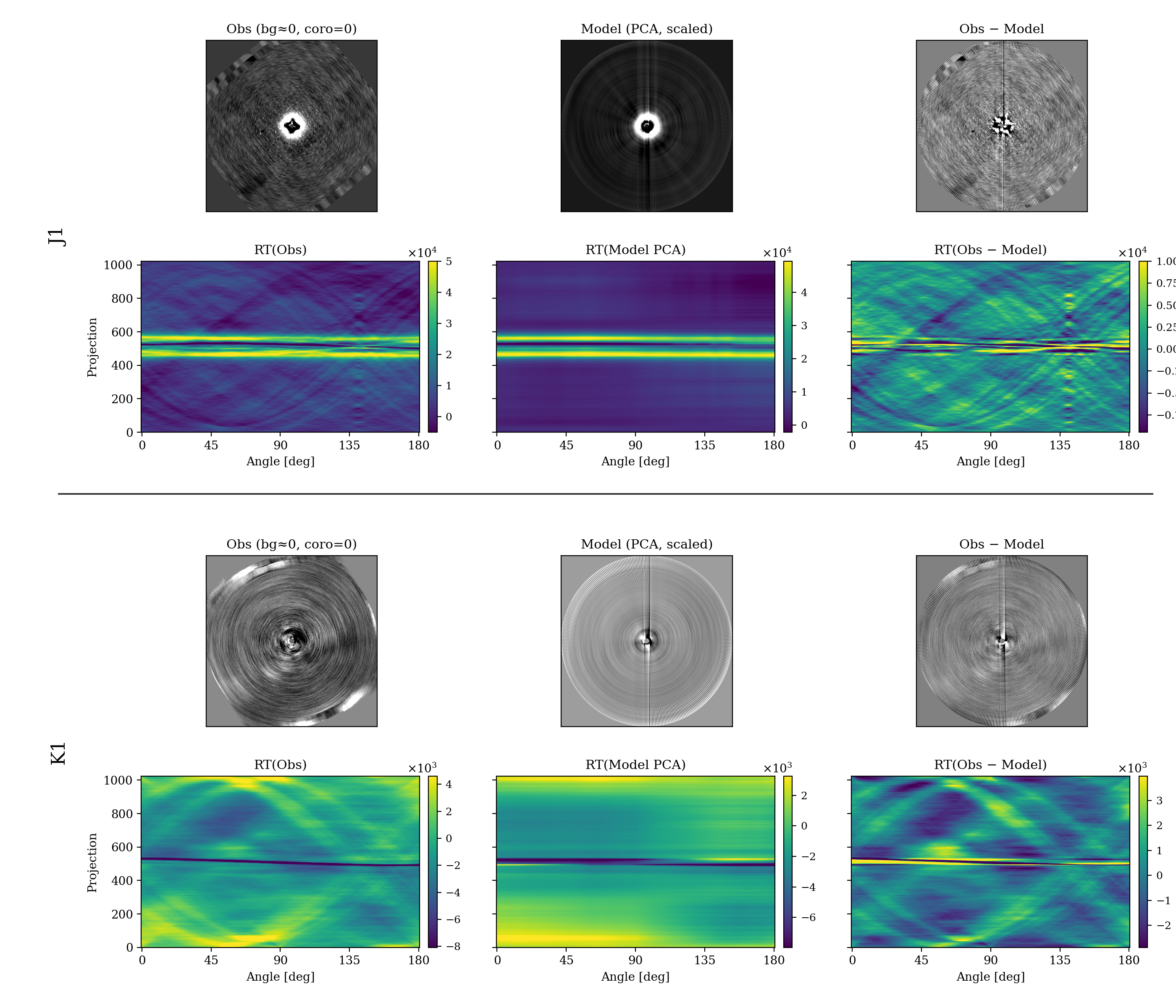}
\figsetgrpnote{Composite PCA-Radon comparison in the J1 (top) and K1 (bottom) filters. In each panel, the top row shows the observed coronagraphic image (left), the model reconstructed from the PCA decomposition in Radon space (middle), and the residual image obtained by inverse Radon transform of the residual sinogram (right). The bottom row shows the Radon sinogram of the observed image, the PCA model sinogram, and the residual sinogram obtained by subtracting the model from the observed data in projection space.}
\figsetgrpend

\figsetgrpstart
\figsetgrpnum{2.6}
\figsetgrptitle{HD 67594}
\figsetplot{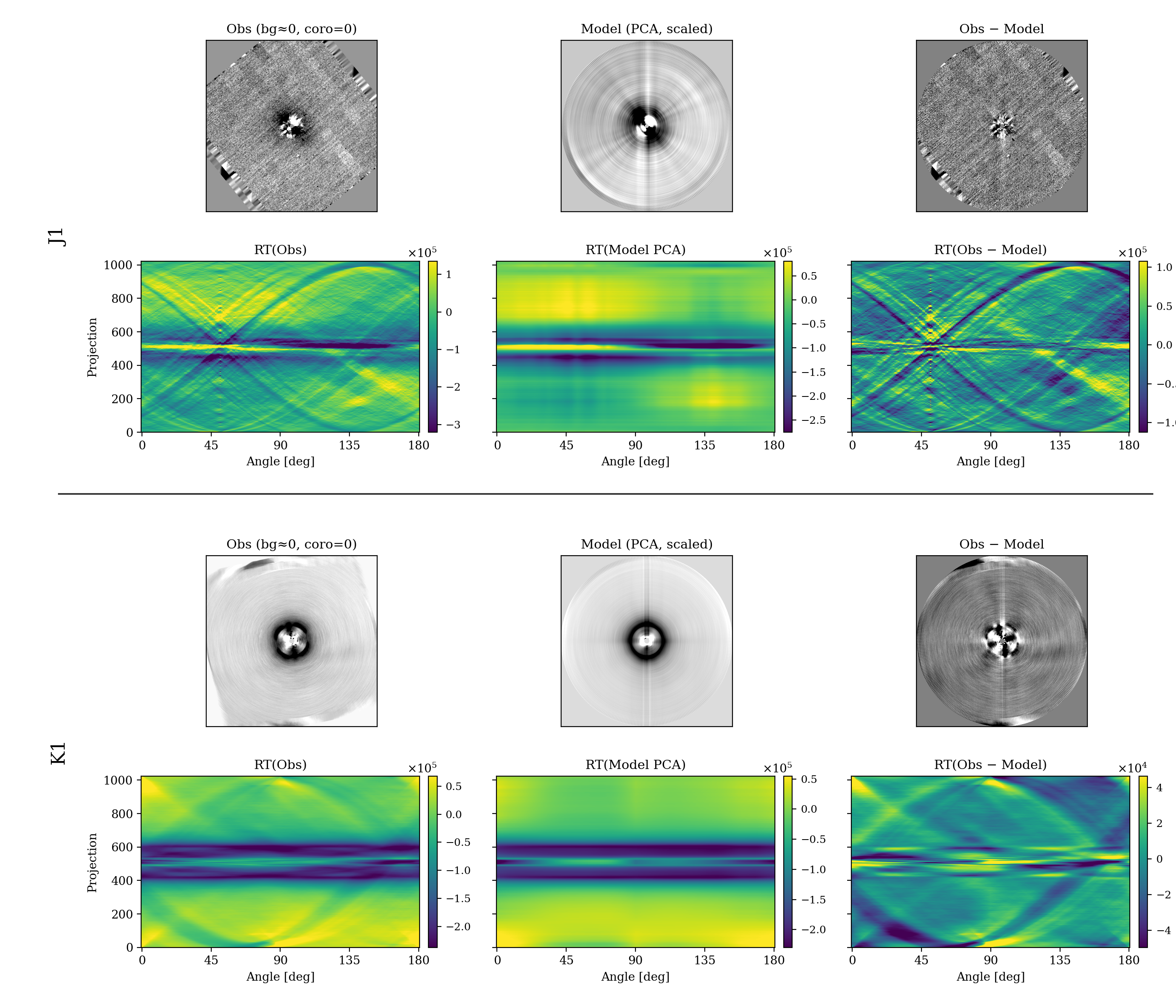}
\figsetgrpnote{Composite PCA-Radon comparison in the J1 (top) and K1 (bottom) filters. In each panel, the top row shows the observed coronagraphic image (left), the model reconstructed from the PCA decomposition in Radon space (middle), and the residual image obtained by inverse Radon transform of the residual sinogram (right). The bottom row shows the Radon sinogram of the observed image, the PCA model sinogram, and the residual sinogram obtained by subtracting the model from the observed data in projection space.}
\figsetgrpend

\figsetgrpstart
\figsetgrpnum{2.7}
\figsetgrptitle{HD 209750}
\figsetplot{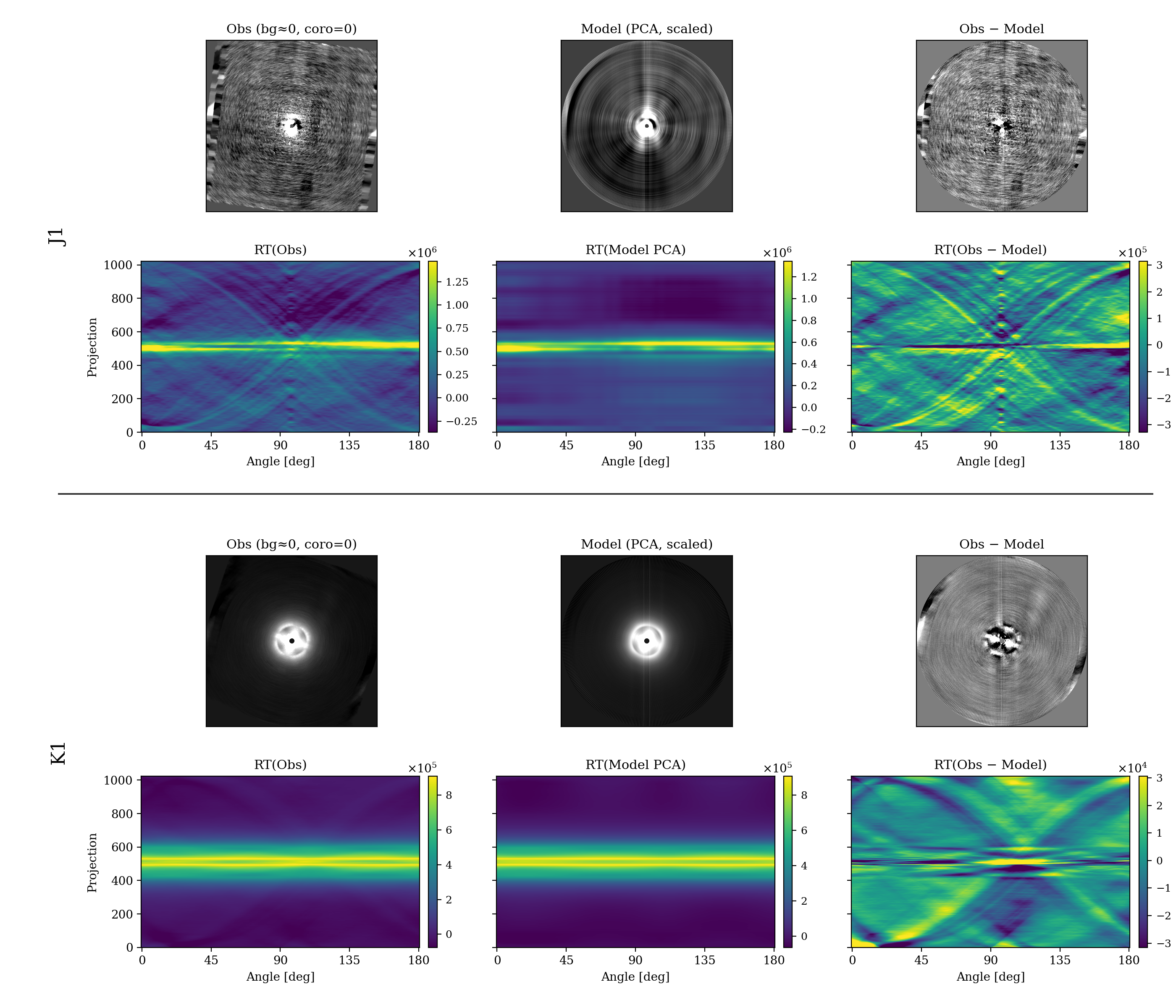}
\figsetgrpnote{Composite PCA-Radon comparison in the J1 (top) and K1 (bottom) filters. In each panel, the top row shows the observed coronagraphic image (left), the model reconstructed from the PCA decomposition in Radon space (middle), and the residual image obtained by inverse Radon transform of the residual sinogram (right). The bottom row shows the Radon sinogram of the observed image, the PCA model sinogram, and the residual sinogram obtained by subtracting the model from the observed data in projection space.}
\figsetgrpend

\figsetend

% ============================================================
% REPRESENTATIVE FIGURE SHOWN IN PDF
% ============================================================

% ============================================================
% ONLINE FIGURE SET 3
% ============================================================

\figsetstart
\figsetnum{3}
\figsettitle{Azimuthally averaged radial intensity profiles}

\figsetgrpstart
\figsetgrpnum{3.1}
\figsetgrptitle{HD 74395}
\figsetplot{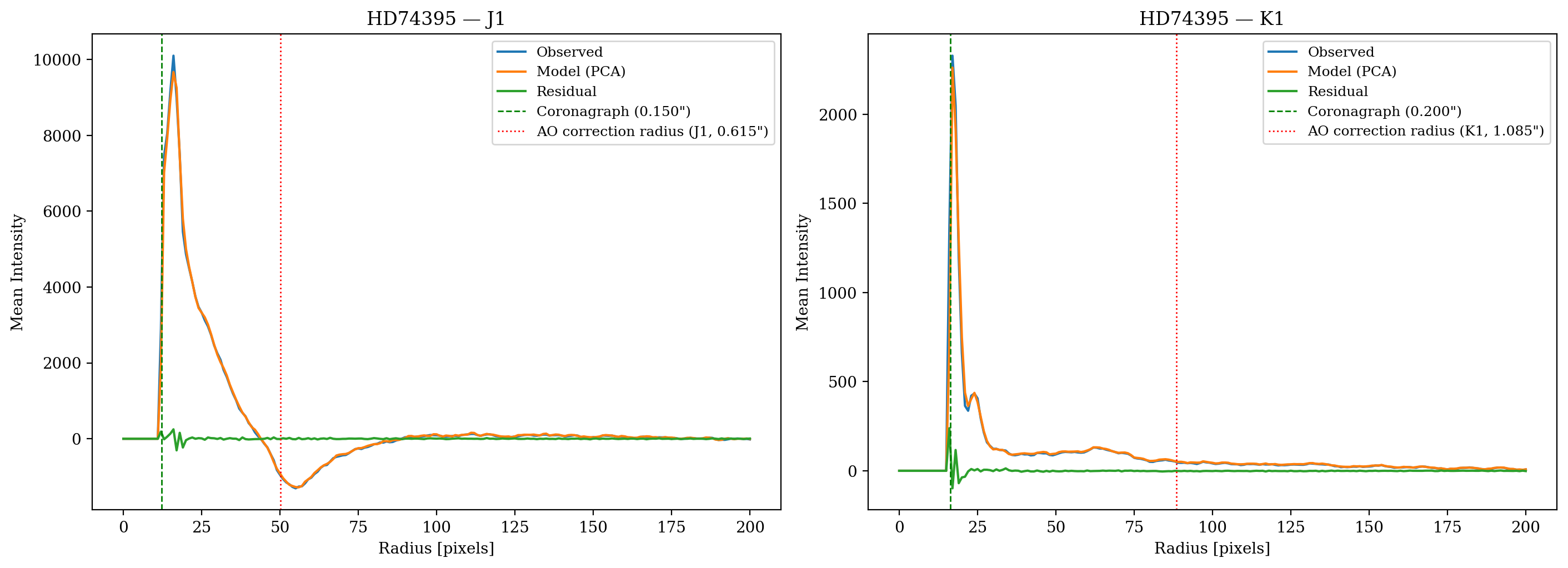}
\figsetgrpnote{Azimuthally averaged radial intensity profiles in the J1 (left) and K1 (right) filters after background subtraction and coronagraph masking. In both panels, the blue curve shows the observed coronagraphic image, the orange curve shows the PCA-based Radon-space model reconstructed using the dominant projection modes, and the green curve shows the residual profile obtained from the difference between the observed and modeled data. The dashed vertical lines mark the coronagraph radii (0.150'' for J1 and 0.200'' for K1), while the dotted vertical lines indicate the AO correction radii (0.615'' for J1 and 1.085'' for K1).}
\figsetgrpend

\figsetgrpstart
\figsetgrpnum{3.2}
\figsetgrptitle{HD 38915}
\figsetplot{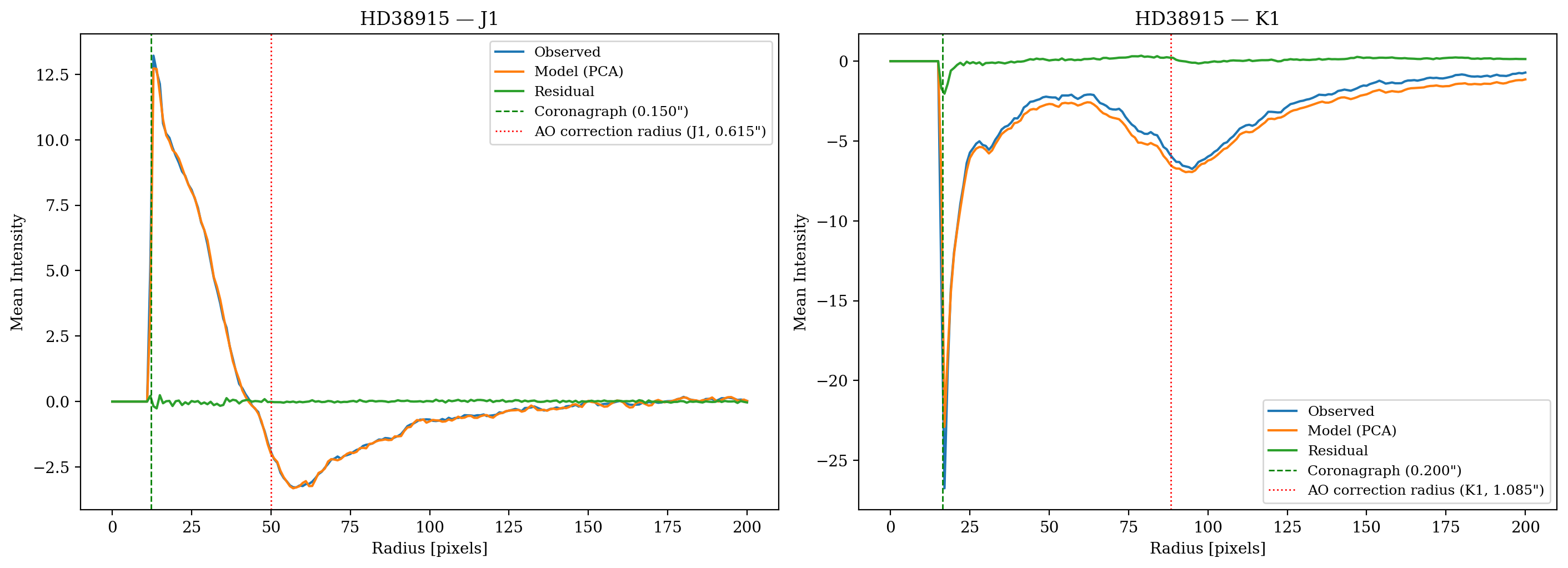}
\figsetgrpnote{Azimuthally averaged radial intensity profiles in the J1 (left) and K1 (right) filters after background subtraction and coronagraph masking. In both panels, the blue curve shows the observed coronagraphic image, the orange curve shows the PCA-based Radon-space model reconstructed using the dominant projection modes, and the green curve shows the residual profile obtained from the difference between the observed and modeled data. The dashed vertical lines mark the coronagraph radii (0.150'' for J1 and 0.200'' for K1), while the dotted vertical lines indicate the AO correction radii (0.615'' for J1 and 1.085'' for K1). The K1 residual profile exhibits a negative baseline offset, likely caused by over-subtraction in the RDI process.}
\figsetgrpend

\figsetgrpstart
\figsetgrpnum{3.3}
\figsetgrptitle{HD 50877}
\figsetplot{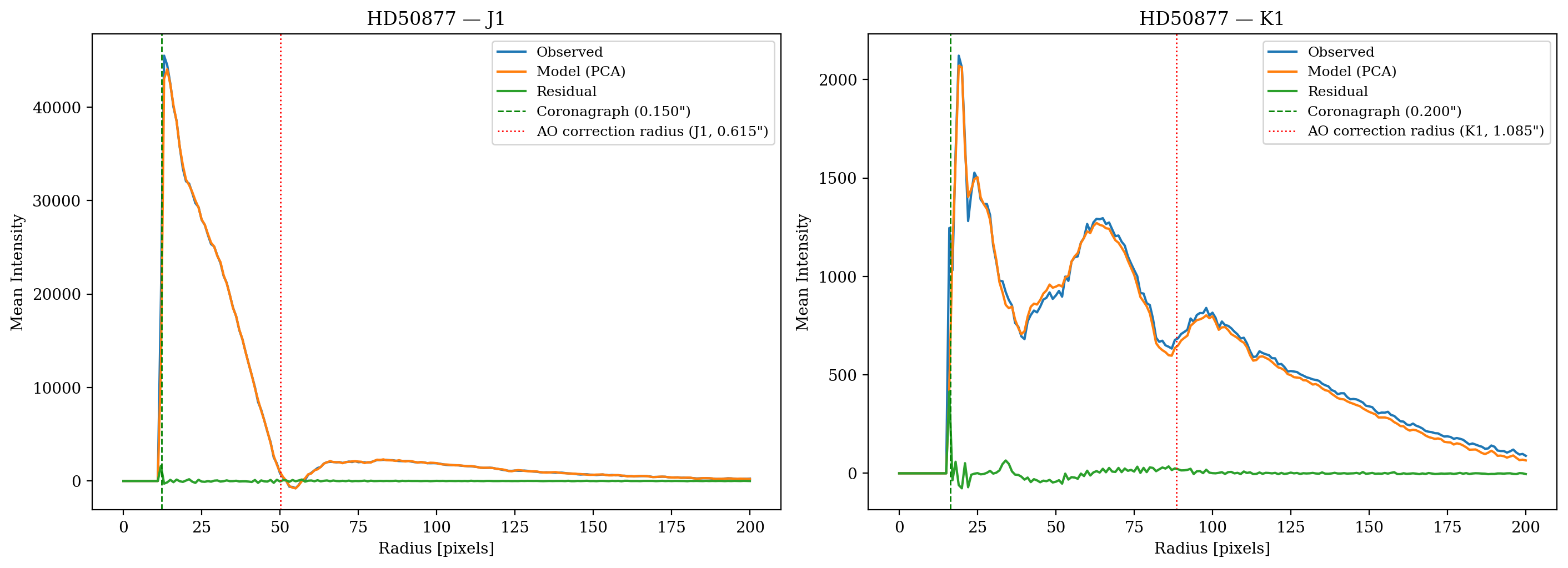}
\figsetgrpnote{Azimuthally averaged radial intensity profiles in the J1 (left) and K1 (right) filters after background subtraction and coronagraph masking. In both panels, the blue curve shows the observed coronagraphic image, the orange curve shows the PCA-based Radon-space model reconstructed using the dominant projection modes, and the green curve shows the residual profile obtained from the difference between the observed and modeled data. The dashed vertical lines mark the coronagraph radii (0.150'' for J1 and 0.200'' for K1), while the dotted vertical lines indicate the AO correction radii (0.615'' for J1 and 1.085'' for K1).}
\figsetgrpend

\figsetgrpstart
\figsetgrpnum{3.4}
\figsetgrptitle{HD 52877}
\figsetplot{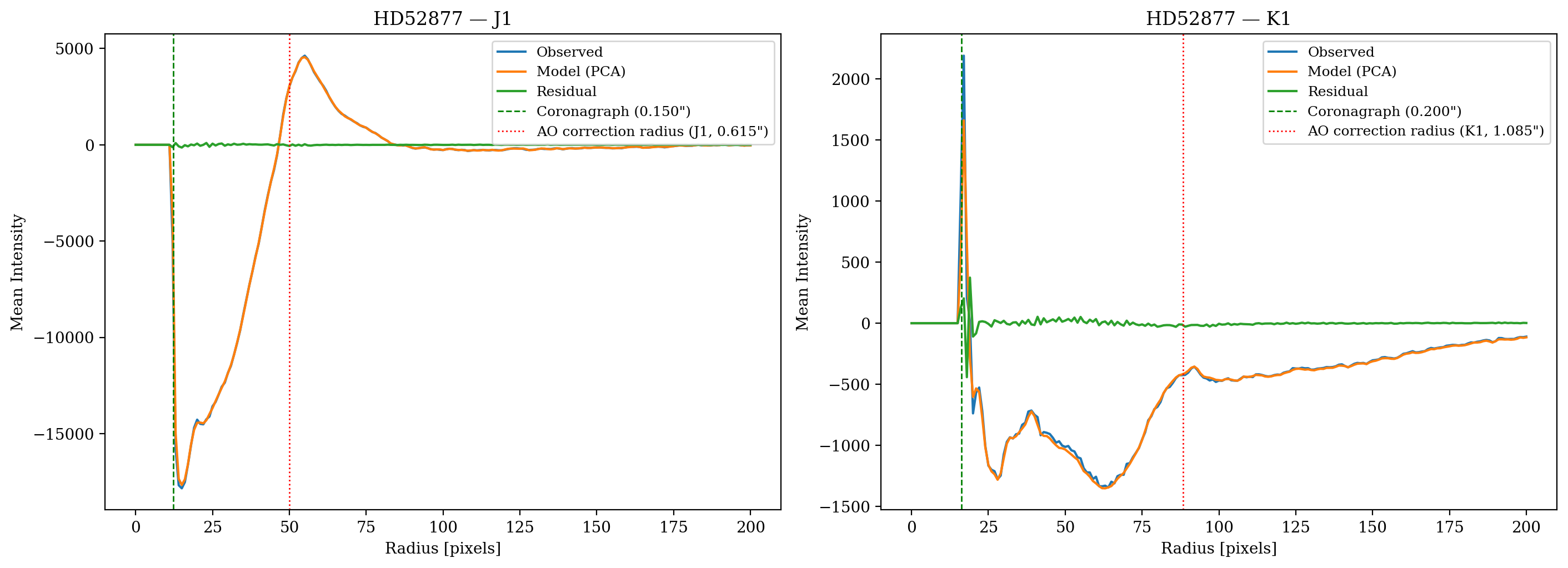}
\figsetgrpnote{Azimuthally averaged radial intensity profiles in the J1 (left) and K1 (right) filters after background subtraction and coronagraph masking. In both panels, the blue curve shows the observed coronagraphic image, the orange curve shows the PCA-based Radon-space model reconstructed using the dominant projection modes, and the green curve shows the residual profile obtained from the difference between the observed and modeled data. The dashed vertical lines mark the coronagraph radii (0.150'' for J1 and 0.200'' for K1), while the dotted vertical lines indicate the AO correction radii (0.615'' for J1 and 1.085'' for K1).}
\figsetgrpend

\figsetgrpstart
\figsetgrpnum{3.5}
\figsetgrptitle{HD 58526}
\figsetplot{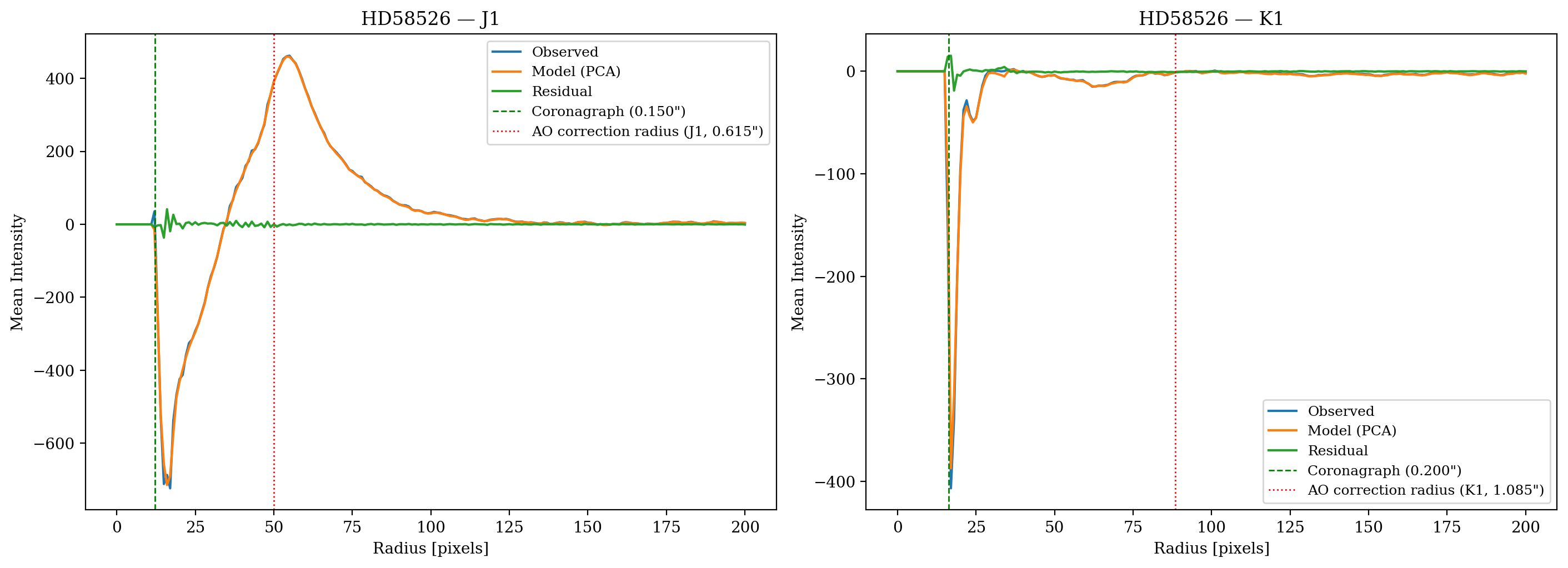}
\figsetgrpnote{Azimuthally averaged radial intensity profiles in the J1 (left) and K1 (right) filters after background subtraction and coronagraph masking. In both panels, the blue curve shows the observed coronagraphic image, the orange curve shows the PCA-based Radon-space model reconstructed using the dominant projection modes, and the green curve shows the residual profile obtained from the difference between the observed and modeled data. The dashed vertical lines mark the coronagraph radii (0.150'' for J1 and 0.200'' for K1), while the dotted vertical lines indicate the AO correction radii (0.615'' for J1 and 1.085'' for K1). Some K-band residual profiles fall below zero, consistent with over-subtraction in the RDI process rather than astrophysical emission.}
\figsetgrpend

\figsetgrpstart
\figsetgrpnum{3.6}
\figsetgrptitle{HD 67594}
\figsetplot{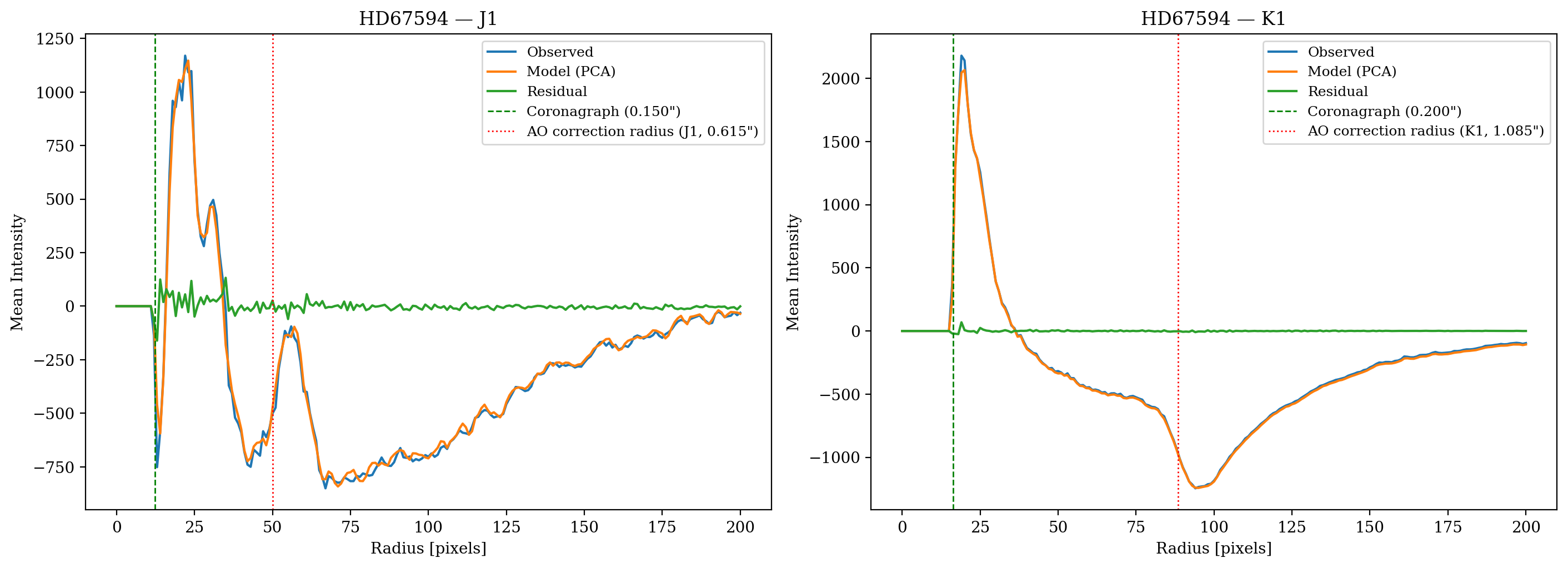}
\figsetgrpnote{Azimuthally averaged radial intensity profiles in the J1 (left) and K1 (right) filters after background subtraction and coronagraph masking. In both panels, the blue curve shows the observed coronagraphic image, the orange curve shows the PCA-based Radon-space model reconstructed using the dominant projection modes, and the green curve shows the residual profile obtained from the difference between the observed and modeled data. The dashed vertical lines mark the coronagraph radii (0.150'' for J1 and 0.200'' for K1), while the dotted vertical lines indicate the AO correction radii (0.615'' for J1 and 1.085'' for K1).}
\figsetgrpend

\figsetgrpstart
\figsetgrpnum{3.7}
\figsetgrptitle{HD 209750}
\figsetplot{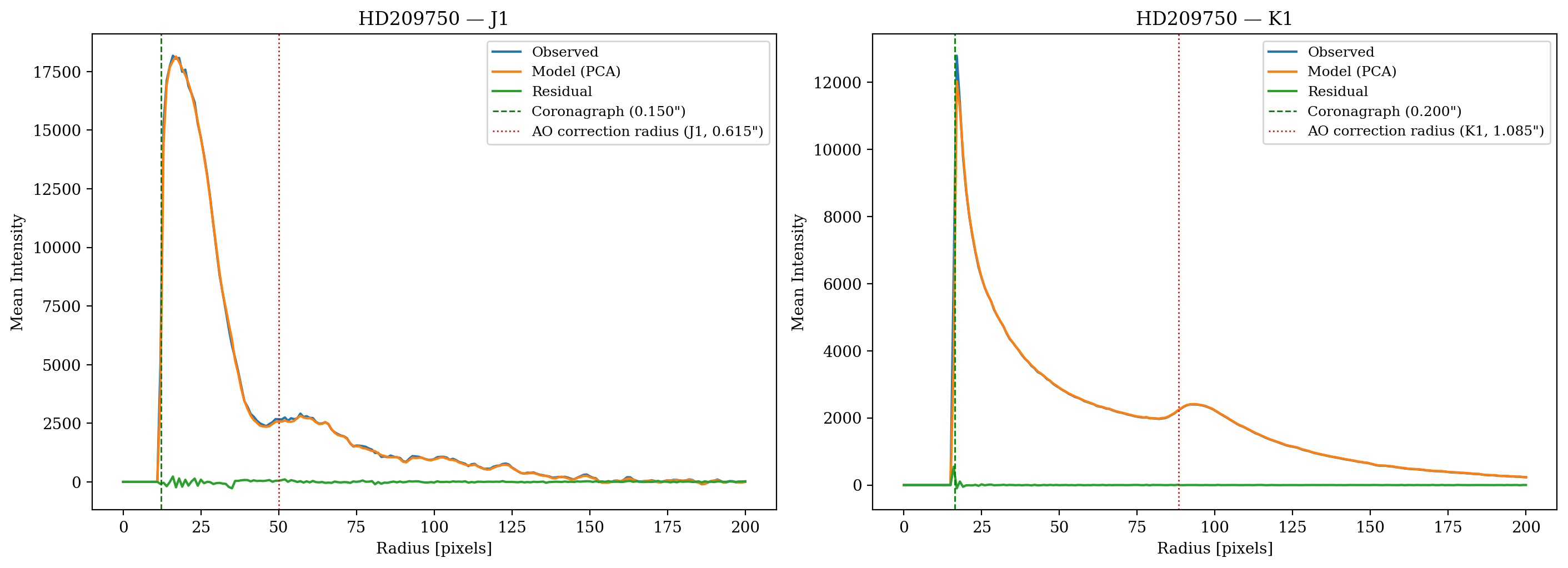}
\figsetgrpnote{Azimuthally averaged radial intensity profiles in the J1 (left) and K1 (right) filters after background subtraction and coronagraph masking. In both panels, the blue curve shows the observed coronagraphic image, the orange curve shows the PCA-based Radon-space model reconstructed using the dominant projection modes, and the green curve shows the residual profile obtained from the difference between the observed and modeled data. The dashed vertical lines mark the coronagraph radii (0.150'' for J1 and 0.200'' for K1), while the dotted vertical lines indicate the AO correction radii (0.615'' for J1 and 1.085'' for K1).}
\figsetgrpend

\figsetend

% ============================================================
% REPRESENTATIVE FIGURE SHOWN IN PDF
% ============================================================

\subsection{HDUST Radiative-transfer Modeling of HD~74395}
\label{Radiative-transfer}

We considered synthetic radiative-transfer models calculated with the 3-dimensional Monte Carlo code \textsc{HDUST} \citep{Carciofi_2006} to investigate whether circumstellar dust around HD~74395 could remain undetected in our SPHERE/IRDIS observations. The models include the stellar photosphere with or without circumstellar dust, with inner dust radii of $R_{\rm in}=15 R_{\star}$ and $80 R_{\star}$. The $R_{\rm in}=15 R_{\star}$ model corresponds approximately to the radius at which the dust temperature reaches the adopted sublimation temperature of $T=1300$~K, while $R_{\rm in}=80 R_\star$ was chosen to represent a configuration without hot inner dust, which is not indicated by the observations. The models have optical depths of $\tau_{\rm V}=0.01$, 0.1, and 1. We adopted $R_{\star}=34.4R_{\odot}$, $d=280$ pc, and $R_{\rm out}=600R_{\star}$. The larger outer radius allows us to test the response of our analysis to circumstellar emission outside the coronagraph and does not assume that the dust physically extends to this radius.

\begin{figure}
    \centering
    \includegraphics[width=0.85\columnwidth]{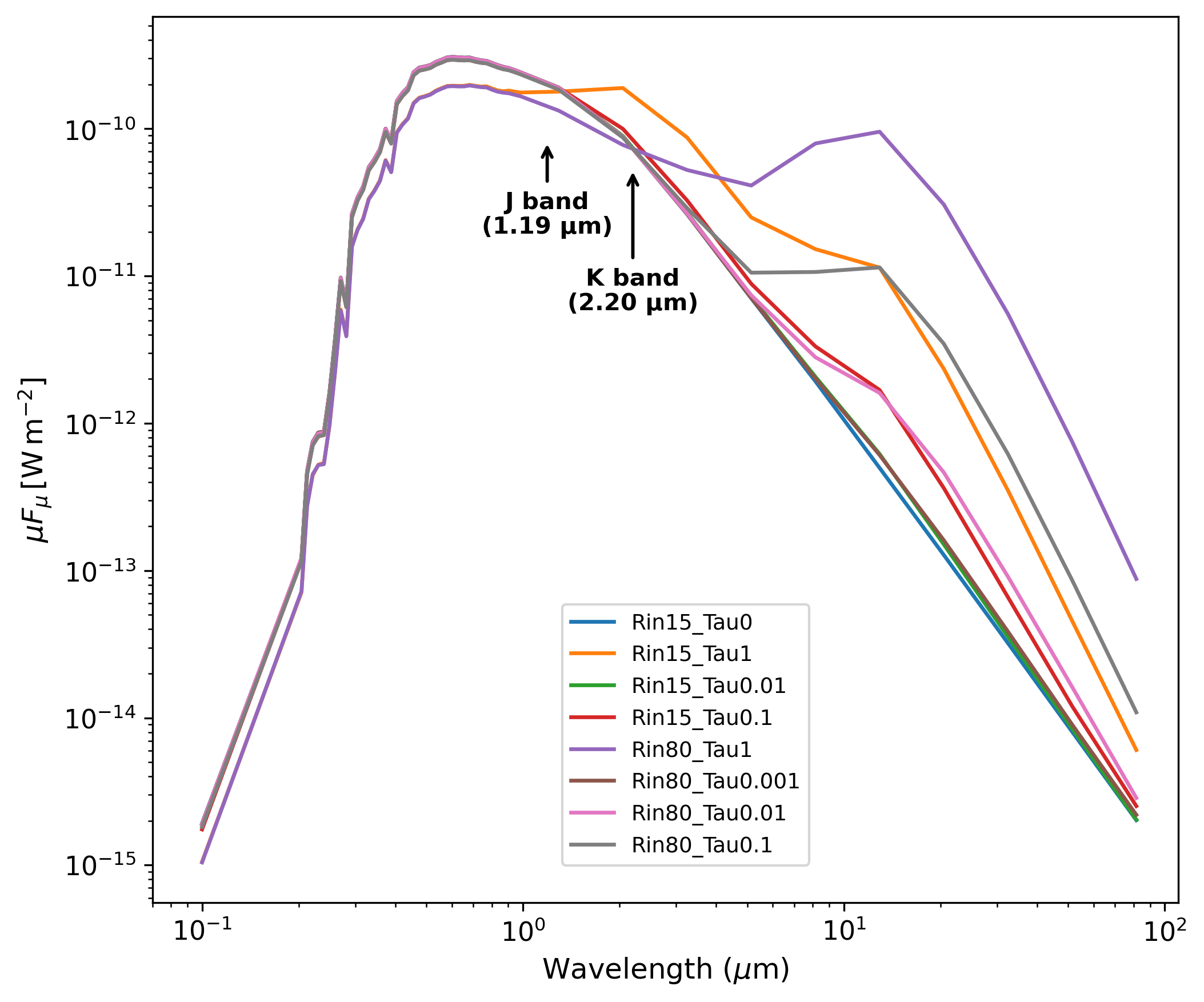}
    \caption{ Synthetic SED for the HD~74395 HDUST models with different inner dust radii and optical depths. The approximate locations of the J and K bands used in the SPHERE/IRDIS observations are indicated by arrows. The differences between the models become more prominent at wavelengths longer than those probed by the J- and K-band observations.}
    \label{fig:hdust_sed}
\end{figure}

Figure~\ref{fig:hdust_sed} shows the synthetic SEDs with and without dust, with the J and K bands indicated. The dusty models show an infrared excess relative to the photosphere-only model, with the strongest excess occurring at wavelengths longer than the SPHERE J and K bands. At $1.19\mu{\rm m}$, the $\tau_{\rm V}=0.01$, 0.1, and 1 models differ from the photosphere-only model by approximately $0.13\%$, $1.26\%$, and $14.2\%$, respectively, for $R_{\rm in}=15R_\star$. We therefore adopted the $\tau_{\rm V}=1$ model for the spatial analysis.

\begin{figure}
    \centering
    \includegraphics[width=\textwidth]{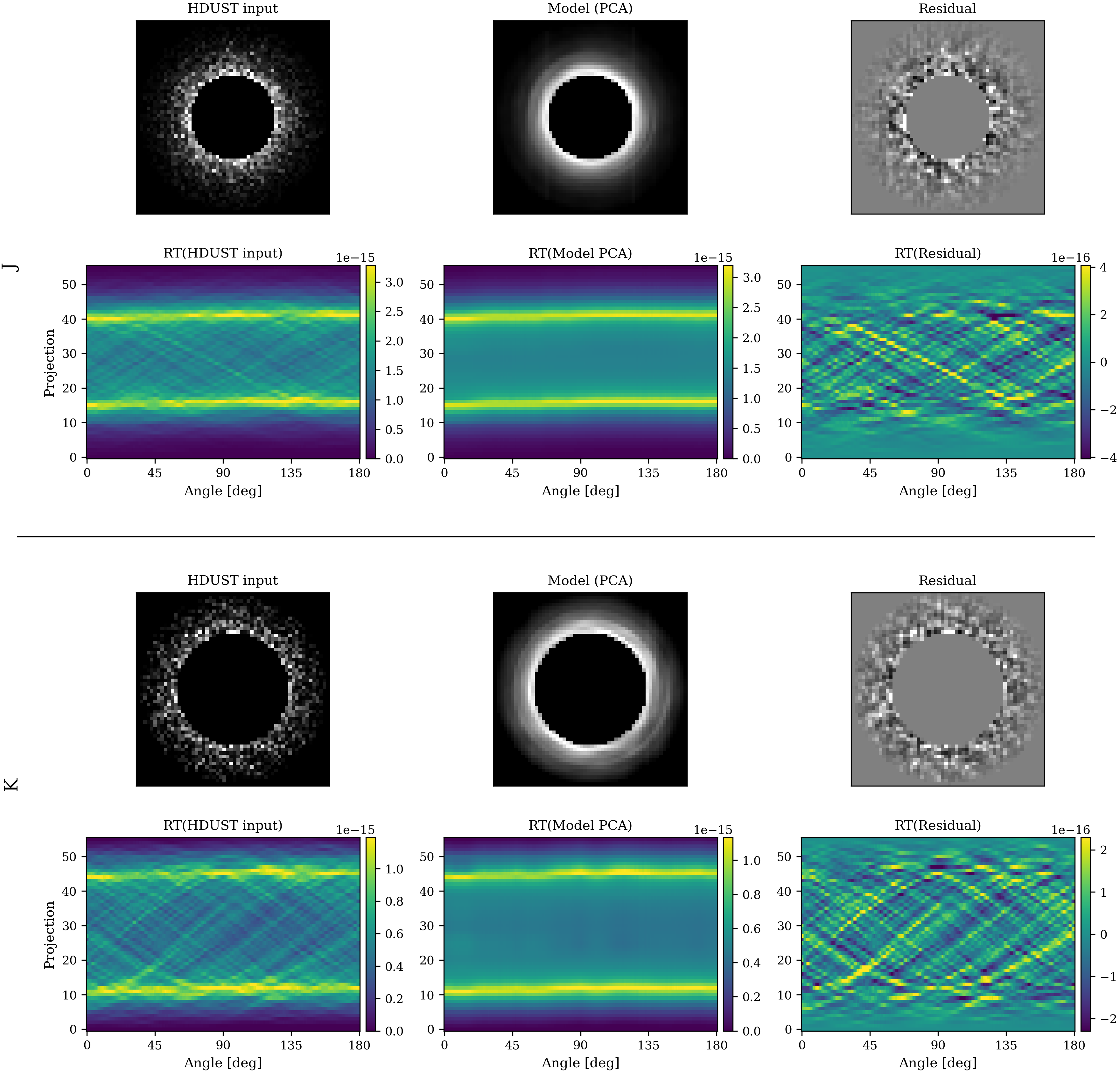}

    \caption{Application of the Radon transform/PCA analysis to the synthetic J-band (top) and K-band (bottom) \textsc{HDUST} models for HD~74395 with $R_{\rm in}=15\,R_\star$, $R_{\rm out}=600\,R_\star$, and $\tau_{\rm V}=1$. The top and bottom sets show the J- and K-band results, respectively. For each band, the upper panels show the coronagraph-masked \textsc{HDUST} image, PCA model, and residual, while the lower panels show their corresponding Radon transforms.}
    \label{fig:hdust_radon}
\end{figure}

\begin{figure}
    \centering
    \includegraphics[width=0.95\columnwidth]{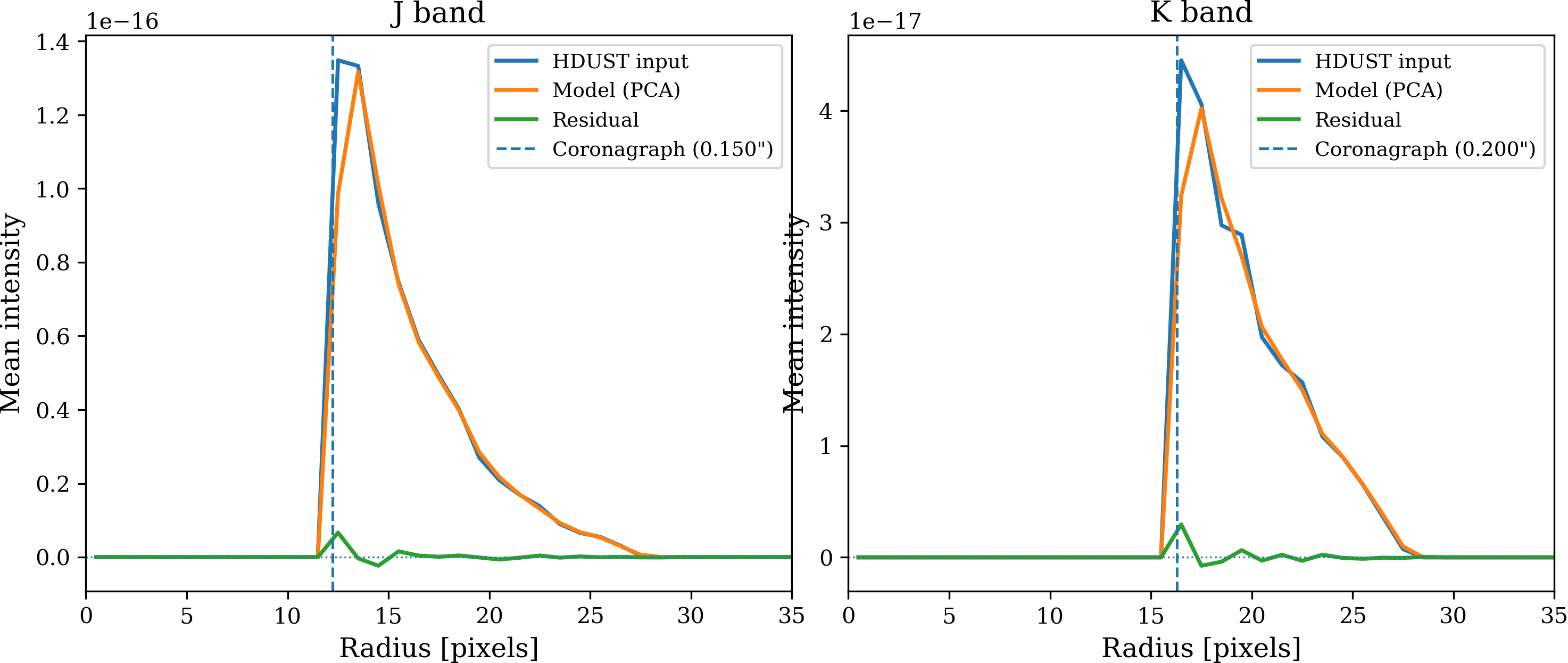}

    \caption{Azimuthally averaged radial intensity profiles for the J-band (left) and K-band (right) \textsc{HDUST} models shown in Figure~\ref{fig:hdust_radon}. The profiles show the input \textsc{HDUST} model, PCA model, and residual after subtraction. The dashed vertical lines mark the adopted coronagraphic radii for each band.}
    \label{fig:hdust_radial}
\end{figure}

We resampled the synthetic images to the angular scale of HD~74395 and the IRDIS plate scale and masked the central region corresponding to the coronagraph. We then analyzed the models following the same Radon transform/PCA procedure used for the observational data (Figure~\ref{fig:hdust_radon}). The broad circumstellar emission is well reproduced by the PCA model, while some angle-dependent structure remains in the residual. The radial intensity profiles (Figure~\ref{fig:hdust_radial}) also show a good fit between the \textsc{HDUST} emission and the PCA model. The largest residuals occur near the coronagraph, with smaller residuals at larger separations.

These models provide a qualitative test rather than a quantitative prediction of detectability because we did not convolve the synthetic images with the instrumental PSF or inject them into the observational noise. The models show that circumstellar material close to the star may remain hidden by the coronagraph, while cooler material at larger radii may emit primarily beyond the J and K bands. Our non-detections therefore do not exclude circumstellar material outside the spatial and wavelength ranges probed by these observations.

\section{What Limits the Detection of Diffuse Circumstellar Emission?}\label{sec:discussion}

\subsection{Why Is Diffuse Circumstellar Material Difficult to Detect?}

Theory predicts massive stars undergo mass-loss throughout their post-main-sequence evolution through steady winds and episodic ejection events. In principle, such processes should populate the circumstellar environment with shells, arcs, or density enhancements \citep{Stothers1996, VanLoon2008, Beasor2022}. However, the absence of detectable diffuse emission in our sample suggests that, for these targets, any circumstellar material must be either intrinsically faint or distributed at very low surface brightness levels.

Independent constraints from other astrophysical probes support this picture. In the context of gamma-ray bursts (GRBs), dense circumstellar structures around massive progenitors are expected to produce observable signatures such as flares or deviations from smooth afterglow light curves, which are often not observed around most evolved stars, suggesting that the surrounding material is of relatively low density \citep{vanMarle2006}.

High-angular-resolution imaging studies have reached similar conclusions and tend to detect circumstellar material primarily in unusually dusty or extreme systems \citep{Kervella2016, Cannon2021, Lomax2025}. Together, these results suggest that while mass ejection mechanisms operate in massive stars, the resulting circumstellar structures are often too diffuse to be readily detected in scattered light.

\subsection{Spectral Energy Distribution of Sample Stars}

Given the absence of detectable circumstellar emission in the high-contrast images, we need to investigate whether this non-detection reflects an intrinsic lack of circumstellar material or observational limitations. To test the former, we examined the spectral energy distributions (SEDs) of our sample stars for evidence of IR excess indicative of diffuse circumstellar material. While SPHERE/IRDIS is sensitive to spatially extended emission at sub-arcsecond scales, SED analysis provides an independent constraint through broadband flux measurements.

We constructed the SEDs of all sample stars using broadband photometric measurements from archival catalogs, including \textit{2MASS}, \textit{WISE}, \textit{IRAS}, \textit{Gaia}, and \textit{Tycho}, obtained through the VizieR Catalogue \citep{Ochsenbein2000}, and converted them to flux density as a function of wavelength. To model the photospheric emission, we fitted a blackbody function using the \texttt{BlackBody} model in the \texttt{astropy.modeling} library \citep{Astropy2018} to the optical–near-IR regime (0.39–2.2~$\mu$m), where the stellar photosphere dominates. We treated both the effective temperature and a normalization factor as free parameters. The normalization accounts for the stellar solid angle and distance. Since we did not explicitly correct the photometry for interstellar extinction, we interpret the fitted blackbody temperature as an effective color temperature rather than a precise literature $T_{\rm eff}$, and we report an uncertainty range for the fitted temperature.

We performed the fit using weighted least squares, with weights based on the reported catalog uncertainties. To ensure that heterogeneous measurements from different surveys did not allow a small subset of points with unusually small quoted uncertainties to dominate the fit, we applied a 5$\sigma$ clipping procedure to the logarithmic residuals within the fitted wavelength range. This step removed inconsistent or spurious catalog measurements while preserving potential long-wavelength excess emission.

\begin{figure}[h]
    \centering
    \includegraphics[width=\textwidth]{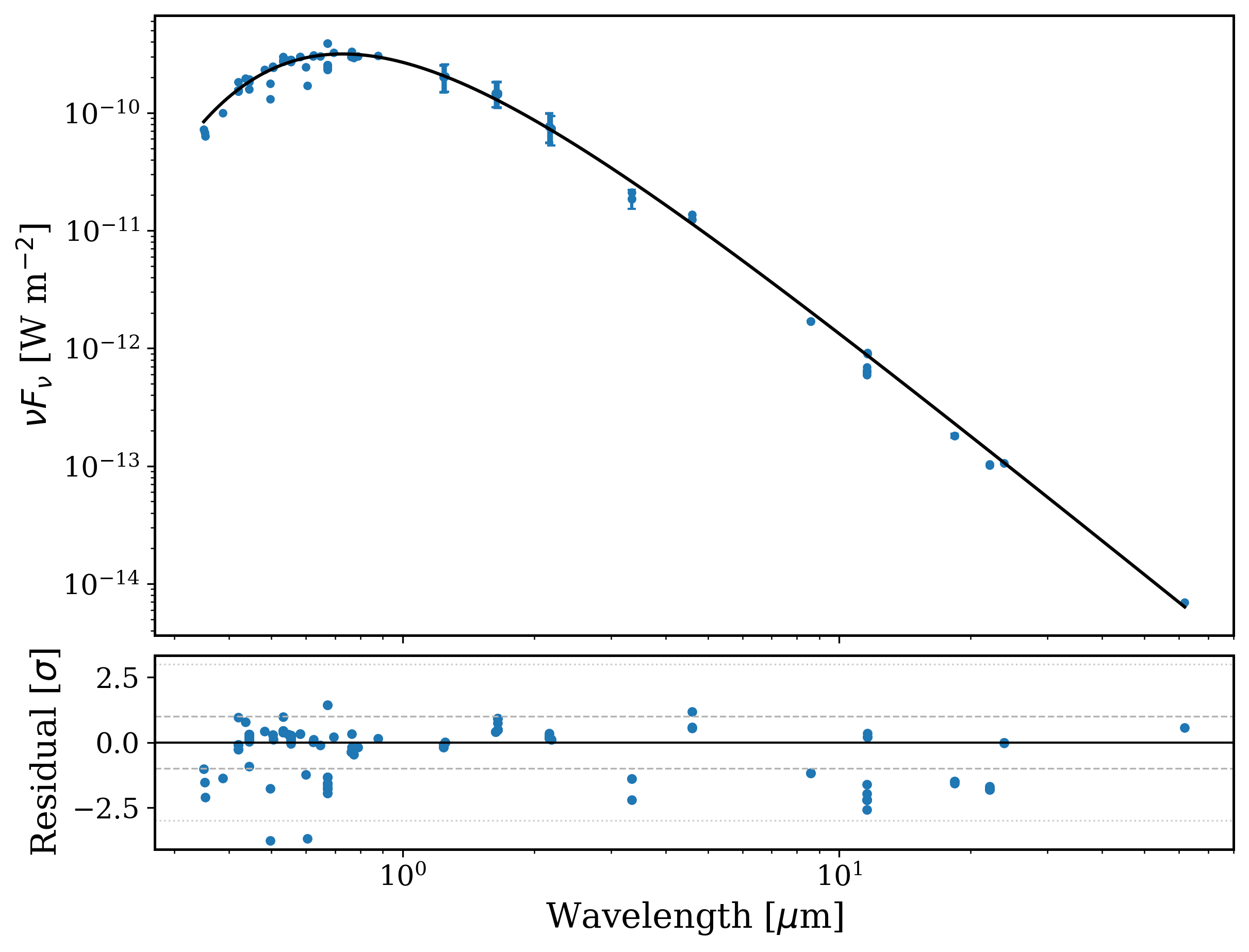}
    \caption{SED of HD~74395 constructed from archival photometry obtained from VizieR, including data from \textit{2MASS}, \textit{WISE}, \textit{IRAS}, \textit{Gaia}, and \textit{Tycho}. Blue points show the catalog photometric measurements. The black line represents the best-fit blackbody model. The model reproduces the optical–near-IR flux distribution, and no significant IR excess is detected.}
    \label{HD74395-SED}
\end{figure}

The derived temperatures are broadly consistent with literature spectral classifications, with modest systematic offsets likely due to the absence of explicit extinction correction and the use of heterogeneous broadband photometry. We find no statistically significant IR excess at wavelengths $\gtrsim 2~\mu$m for any star in the sample. Figure~\ref{HD74395-SED} shows the SED of HD~74395 as a representative example, with the SEDs of the remaining targets provided in the accompanying online figure set. The Appendix lists the best-fit blackbody temperatures and $\chi^2$ values. For comparison, dusty systems such as WR~140 exhibit pronounced IR excess at wavelengths $\gtrsim 2~\mu$m due to thermal emission from circumstellar dust \citep{Lau2023}. In contrast, the SEDs of our sample are consistent with predominantly photospheric emission. This absence of IR excess is consistent with the non-detection of extended emission in the high-contrast images.

% ============================================================
% ONLINE FIGURE SET 3
% ============================================================
\figsetstart
\figsetnum{8}
\figsettitle{Spectral energy distributions of the sample stars}

\figsetgrpstart
\figsetgrpnum{8.1}
\figsetgrptitle{HD 74395}
\figsetplot{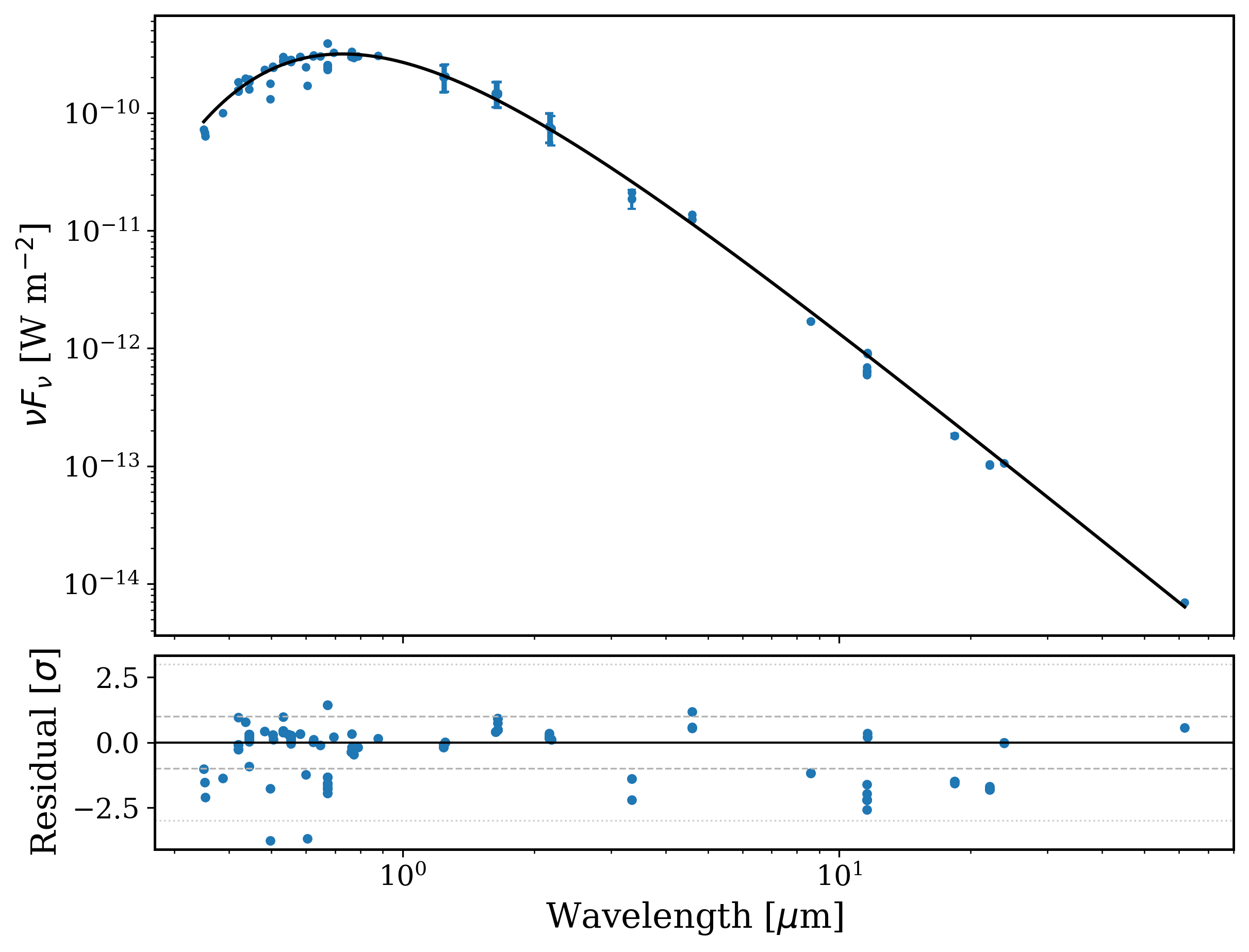}
\figsetgrpnote{SED constructed from archival photometry obtained from VizieR, including data from 2MASS, WISE, IRAS, Gaia, and Tycho. Blue points show the catalog photometric measurements. The black line represents the best-fit blackbody model. The model reproduces the optical–near-IR flux distribution, and no significant IR excess is detected.}
\figsetgrpend

\figsetgrpstart
\figsetgrpnum{8.2}
\figsetgrptitle{HD 38915}
\figsetplot{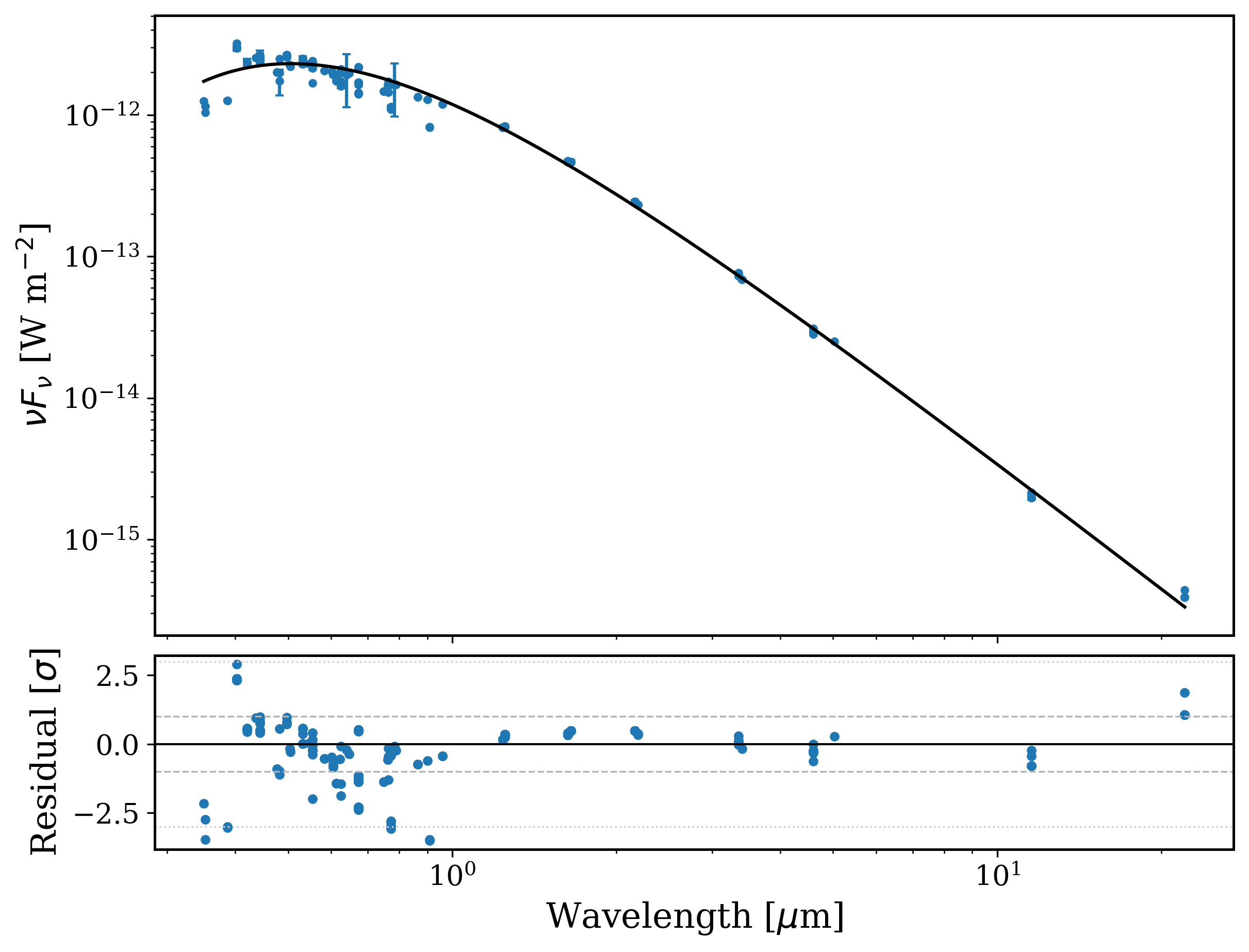}
\figsetgrpnote{SED constructed from archival photometry obtained from VizieR, including data from 2MASS, WISE, IRAS, Gaia, and Tycho. Blue points show the catalog photometric measurements. The black line represents the best-fit blackbody model. The model reproduces the optical–near-IR flux distribution, and no significant IR excess is detected.}
\figsetgrpend

\figsetgrpstart
\figsetgrpnum{8.3}
\figsetgrptitle{HD 50877}
\figsetplot{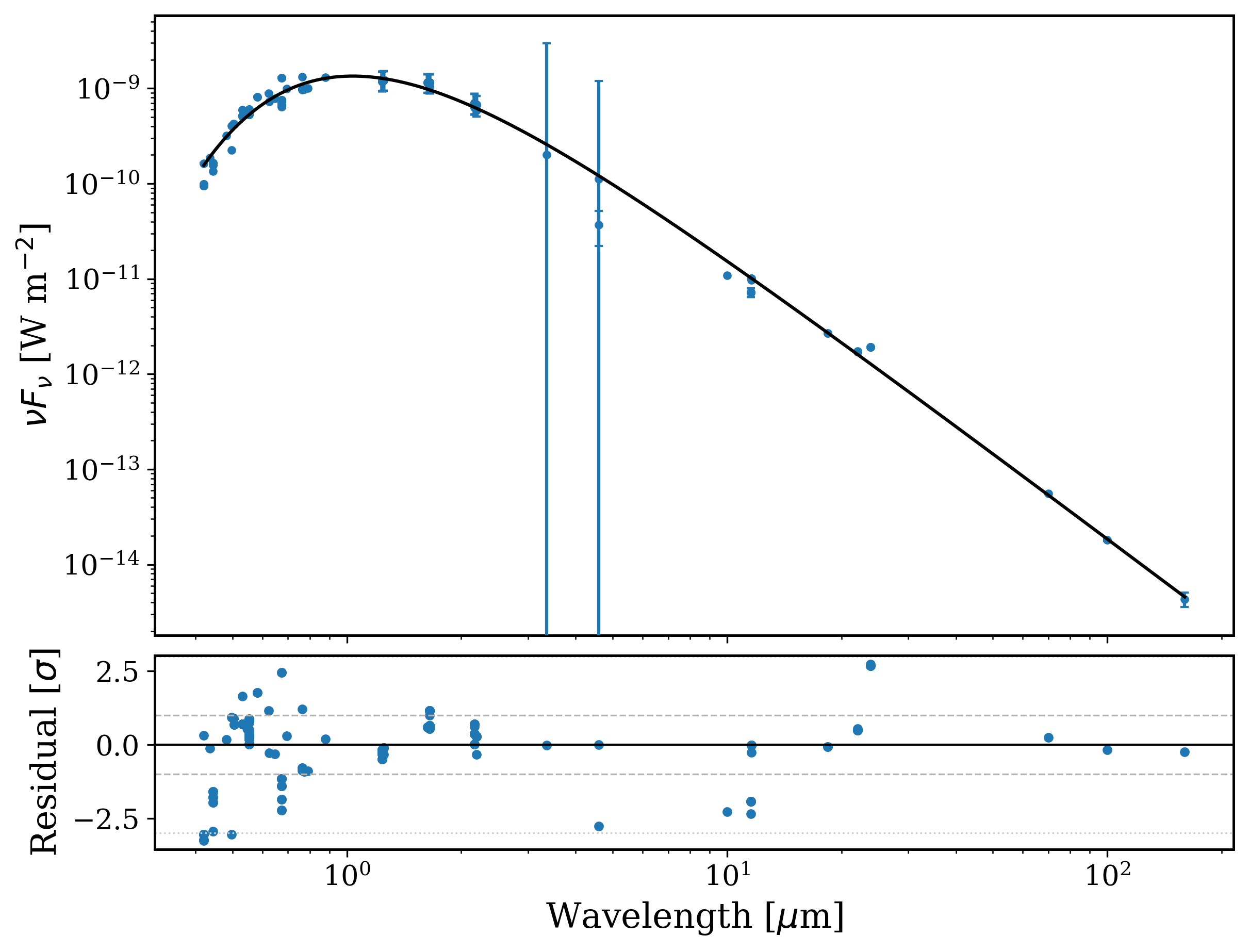}
\figsetgrpnote{SED constructed from archival photometry obtained from VizieR, including data from 2MASS, WISE, IRAS, Gaia, and Tycho. Blue points show the catalog photometric measurements. The black line represents the best-fit blackbody model. The model reproduces the optical–near-IR flux distribution, and no significant IR excess is detected.}
\figsetgrpend

\figsetgrpstart
\figsetgrpnum{8.4}
\figsetgrptitle{HD 52877}
\figsetplot{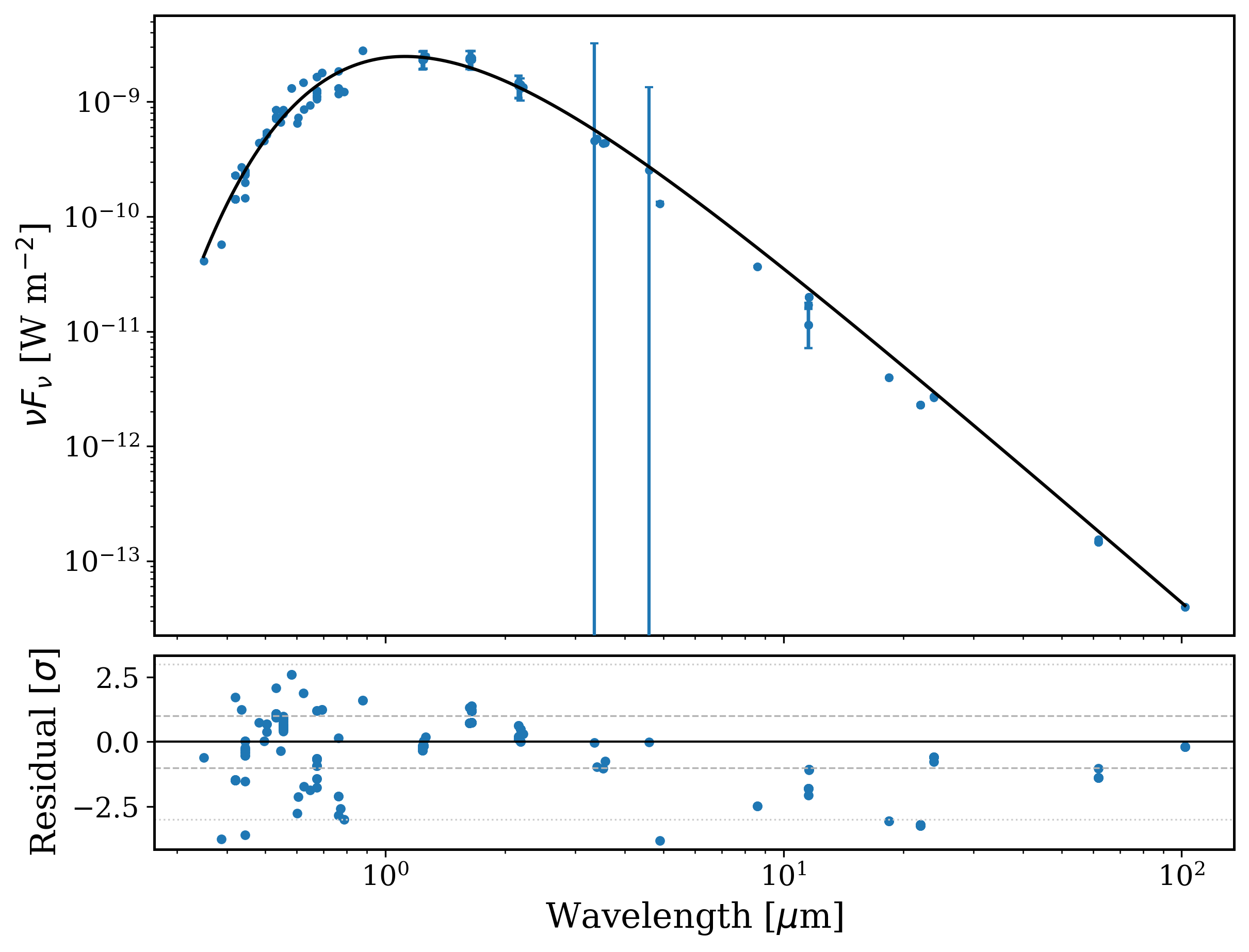}
\figsetgrpnote{SED constructed from archival photometry obtained from VizieR, including data from 2MASS, WISE, IRAS, Gaia, and Tycho. Blue points show the catalog photometric measurements. The black line represents the best-fit blackbody model. The model reproduces the optical–near-IR flux distribution, and no significant IR excess is detected.}
\figsetgrpend

\figsetgrpstart
\figsetgrpnum{8.5}
\figsetgrptitle{HD 58526}
\figsetplot{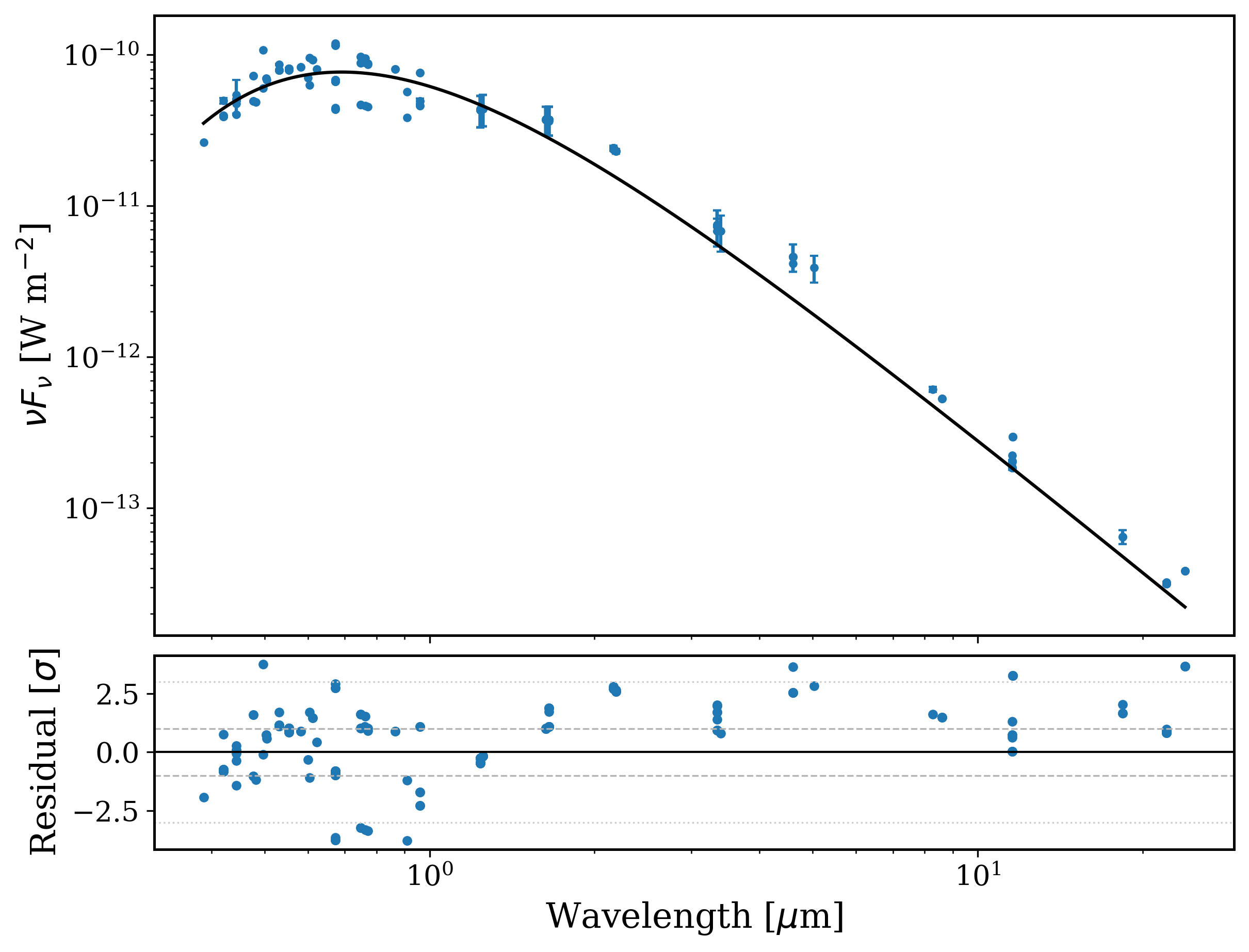}
\figsetgrpnote{SED constructed from archival photometry obtained from VizieR, including data from 2MASS, WISE, IRAS, Gaia, and Tycho. Blue points show the catalog photometric measurements. The black line represents the best-fit blackbody model. The model reproduces the optical–near-IR flux distribution, and no significant IR excess is detected.}
\figsetgrpend

\figsetgrpstart
\figsetgrpnum{8.6}
\figsetgrptitle{HD 67594}
\figsetplot{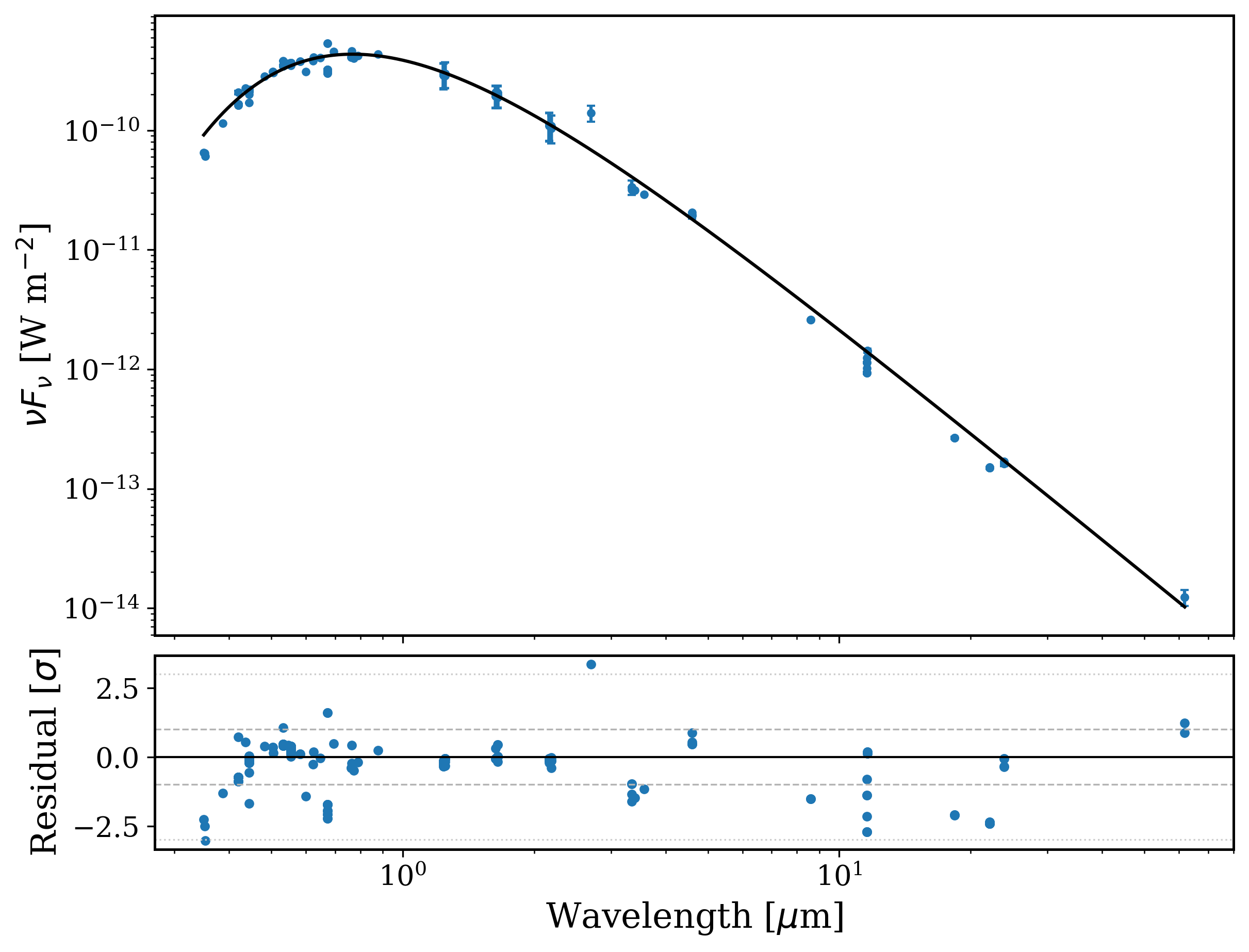}
\figsetgrpnote{SED constructed from archival photometry obtained from VizieR, including data from 2MASS, WISE, IRAS, Gaia, and Tycho. Blue points show the catalog photometric measurements. The black line represents the best-fit blackbody model. The model reproduces the optical–near-IR flux distribution, and no significant IR excess is detected.}
\figsetgrpend

\figsetgrpstart
\figsetgrpnum{8.7}
\figsetgrptitle{HD 209750}
\figsetplot{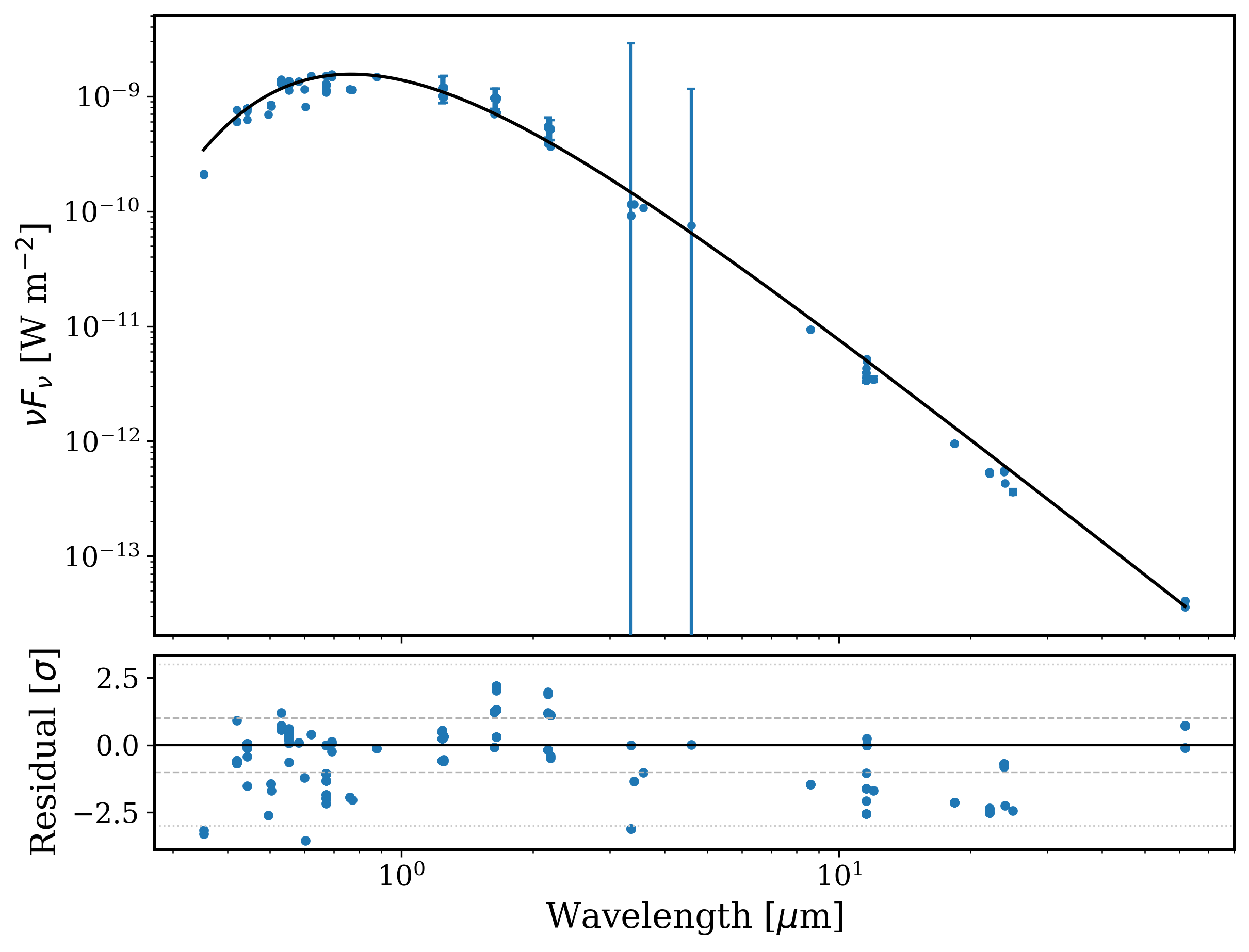}
\figsetgrpnote{SED constructed from archival photometry obtained from VizieR, including data from 2MASS, WISE, IRAS, Gaia, and Tycho. Blue points show the catalog photometric measurements. The black line represents the best-fit blackbody model. The model reproduces the optical–near-IR flux distribution, and no significant IR excess is detected.}
\figsetgrpend

\figsetend

% ============================================================
% REPRESENTATIVE FIGURE SHOWN IN PDF
% ============================================================

\subsection{Improving the Observational Design}

Although the SEDs suggest an intrinsic lack of substantial circumstellar material, we cannot rule out the presence of very low surface brightness emission. We therefore examine the instrumental and observational limitations that may affect our sensitivity. We discuss three primary observational constraints below: 1) exposure time and ND filter limitations, 2) the absence of dedicated sky calibration frames, and 3) the lack of well-matched reference stars for RDI. Together, these factors limit the achievable contrast, particularly at small angular separations.

\subsubsection{Neutral Density Filters and Exposure time limitation}

Accurate absolute flux measurements require an unsaturated PSF, as flux estimation involves matching the counts in the circumstellar region with those from the unsaturated central star. However, the observed PSF is saturated presenting significant challenges in determining the absolute fluxes.

The SPHERE Exposure Time Calculator estimated an ideal DIT that is approximately ten times shorter than SPHERE’s minimum exposure time of 0.83 s. Because the minimum DIT was fixed, avoiding saturation required the use of ND filters. We therefore used ND1 and ND2 filters to attenuate the stellar flux. While these filters reduced the central star's brightness, they also attenuated the faint diffuse circumstellar emission and degraded the SNR in the background. Because the circumstellar emission arises from scattered starlight, reducing the stellar flux proportionally reduces the brightness of this emission, further limiting detectability.

For the present dataset, neither ND1 nor ND2 fully prevented saturation of the brightest targets. In several cases, the central star exceeded 60,000 counts, above the nominal SPHERE/IRDIS saturation limit of $\sim$40,000 counts. At the same time, the attenuation further reduced sensitivity to low surface brightness extended emission. Given the fixed minimum DIT and the brightness of the targets (I $\approx$ 2–9 mag), these instrumental constraints limited our ability to detect faint diffuse circumstellar material.

Future observations with instruments that provide shorter exposure times, increased dynamic range, or improved detector technologies (e.g., next-generation CMOS detectors) could mitigate these limitations. Alternatively, targeting less luminous stars within the dynamic range of SPHERE may improve sensitivity to extended emission.

These combined effects set a practical sensitivity limit for detecting diffuse circumstellar emission in our data. Based on injection–recovery tests for HD 74395, we estimate that structures fainter than $\Delta m \sim 3.5$ mag in J1 and $\Delta m \sim 2.0$ mag in K1  relative to the residual coronagraphic PSF halo are not reliably detectable. These limit reflects the combined impact of instrumental noise, residual halo structure, and the attenuation introduced by ND filters.

\subsubsection{Sky Calibration and Background Stability}

The biggest challenge in working with this data set was the poor quality  SNR in both J and K bands. The raw data were intrinsically noisy, compromising image quality. The K-band data are significantly affected by background residuals due to the absence of dedicated sky calibration frames during the observations. As a result, the background subtraction relies on daytime dark frames, which do not accurately reproduce the on-sky thermal emission. This mismatch introduces additional noise and limits our sensitivity to faint, spatially extended emission.

This limitation is particularly important for studies of spatially extended astrophysical features, as improper background subtraction can either distort or suppress low surface brightness signals. In our dataset, the residual background structure contributes significantly to the overall noise floor, especially in K band, reducing our ability to detect diffuse circumstellar material.

The observing program did not include dedicated sky exposures within the same observing blocks as the K-band observations. In their absence, the standard reduction pipeline used daytime dark-background frames. We did not independently search for suitable sky frames from other archival programs or re-reduce the observations using such frames. Archival sky frames would need to match the DIT, filter configuration, instrumental setup, and observing conditions of the science observations to reproduce the spatial and temporal variability of the thermal background, including contributions from the telescope and instrument optics. Daytime dark-background frames do not fully reproduce the on-sky thermal background, which limits the accuracy of the K-band background subtraction. A library-based approach, such as masked PCA applied to suitable archival sky frames obtained under similar conditions, could improve background modeling in future studies of extended emission.

\subsubsection{Reference star selection for RDI}

For the RDI process, it is crucial to observe an external reference star under the same conditions as the target. Ideally, the observation sequence includes non-coronagraphic images for photometric calibration, coronagraphic observations of the science target, and coronagraphic observations of a reference star of similar spectral type and color obtained within the same observing block.

In our case, suitable reference star observations were not available. We therefore used another of our own targets with similar spectral type and color as a substitute reference for the RDI subtraction. However, this reference star was observed on a different night and under different observing conditions, which leads to mismatches in the PSF structure between the science and reference frames, reducing the effectiveness of the RDI subtraction. This emphasizes the need for future observations to include dedicated reference stars observed under conditions similar to those of the science targets, enabling more effective PSF subtraction and improving sensitivity to faint circumstellar emission.

In addition to improving the RDI observing strategy, future studies could use optical polarimetric imaging to probe circumstellar material at smaller angular separations than are accessible with near-IR observations. Instruments such as SPHERE/ZIMPOL provide higher angular resolution at optical wavelengths, while non-coronagraphic polarimetric observations can probe smaller angular separations and suppress much of the unpolarized stellar light. Such observations could reveal polarized scattered light from circumstellar dust at separations that are inaccessible to our coronagraphic IRDIS observations. However, the detectability of polarized scattered light depends on the dust properties, scattering geometry, and viewing angle. Optical polarimetry therefore provides a complementary approach to the near-infrared total-intensity observations presented here.

\section{Summary and Future Work}
Using a sample of seven nearby Red and Yellow Supergiants, we applied high-contrast differential imaging and Radon transform analysis to search for faint extended emission. Within the sensitivity limits of our data, none of the targets show clear evidence of diffuse circumstellar emission. Based on injection-recovery tests for HD~74395, we recover structures down to contrasts of approximately $\Delta m \sim 3.5$ mag in J1 and $\Delta m \sim 2.0$ mag in K1 relative to the residual coronagraphic PSF halo.

None of the seven stars showed IR excess in their SEDs. However, the HDUST models also show that dust produces more prominent IR excess at longer wavelengths, particularly at higher optical depth. Thus, our J- and K-band observations do not rule out circumstellar material at all radii, including cooler material that may emit more strongly at longer IR wavelengths. 

% Our observations constrain hot diffuse circumstellar emission within the spatial and wavelength ranges accessible to SPHERE/IRDIS. Circumstellar material contributing at J and K may lie predominantly within or close to the coronagraphic region, limiting its detectability in our observations, while cooler material at larger separations may emit more strongly at longer infrared wavelengths.

At the same time, our analysis highlights several observational limitations that affect the achievable sensitivity. In particular, the lack of well-matched reference stars and dedicated sky calibration frames limits the effectiveness of the RDI subtraction. Future observations should include dedicated reference stars matched as closely as possible to the science targets in spectral type, color, brightness, and observing conditions. Such observations would improve PSF subtraction and increase sensitivity to faint circumstellar emission. In addition, the minimum exposure time available with SPHERE/IRDIS (0.83 s), combined with the need to use ND filters for very bright stars, significantly reduces throughput and limits sensitivity close to the coronagraphic mask. We emphasize that coronagraphy remains useful for suppressing stellar light and detecting faint extended emission outside the inner working angle. Even though the coronagraphic mask prevents us from probing the innermost circumstellar regions, removing the coronagraph would increase the stellar contaminations and, in our case, saturate the area around the central star. Since our sample consists of nearby bright supergiants, observing more distant targets would increase the physical scale corresponding to the same coronagraphic inner working angle, further limiting our ability to probe circumstellar material close to the stars where we would expect to detect circumstellar emission at these wavelengths. In addition, our J- and K-band observations probe only near-IR emission, while cooler circumstellar dust may emit more strongly at longer IR wavelengths, as shown by our HDUST models. These spatial and wavelength limitations indicate that, despite its theoretical advantages over other adaptive optics systems, the SPHERE/IRDIS J- and K-band configuration used in this study may not be optimal for detecting very faint, diffuse circumstellar emission around bright evolved stars. A more suitable system would combine the high-contrast performance of SPHERE with detectors that offer shorter readout times and larger well depths and at longer wavelengths, enabling shorter exposures and improved dynamic range.

We recommend that future studies prioritize stars with known or suspected IR excess as targets for detecting diffuse circumstellar emission, while also optimizing observing strategies through dedicated reference stars and improved calibration sequences. Where possible, instruments that allow shorter effective exposure times or higher dynamic range without ND filters may further improve sensitivity to low–surface-brightness emission. Together, these refinements will help better define the conditions under which high-contrast imaging can successfully probe circumstellar material around evolved massive stars. This pilot study therefore provides a realistic benchmark for both the astrophysical and observational constraints that govern the detectability of diffuse circumstellar emission in SPHERE/IRDIS data.

\section*{Acknowledgement}

This research used processed data obtained from the ESO Science Archive Facility with DOI(s): https://doi.org/10.18727/archive/79. The data were originally collected with the SPHERE instrument at the Very Large Telescope under ESO programme 098.D-0460(A). We acknowledge the European Southern Observatory (ESO) staff and the SPHERE instrument team for their efforts in obtaining and maintaining the observations used in this study. This research has made use of the VizieR catalogue access tool, CDS, Strasbourg, France \citep{Ochsenbein2000}. The original description of the VizieR service was published in \citet{Ochsenbein2000}. This research had made use of the SIMBAD database (operated at CDS, Strasbourg, France; \citep{Wenger2000}). P.G., B.G., and N.B.O. acknowledges support in part from NASA EPSCoR 80NNSC22M0063, NSF AST 2319415, and NASA EPSCoR 80NSSC24M0112.

We thank Johan Olofsson for his assessment of the SPHERE data quality and for discussions regarding saturation effects and observational limitations. We thank Jason Aufdenberg and Tomomi Otani for feedback on the manuscript.

% This research used processed data obtained from the ESO Science Archive Facility with DOI(s): https://doi.org/10.18727/archive/79. The data were originally collected with the SPHERE instrument at the Very Large Telescope under ESO programme 098.D-0460(A). We acknowledge the European Southern Observatory (ESO) staff and the SPHERE instrument team for their efforts in obtaining and maintaining the observations used in this study. Data reduction/archival products were provided by the High Contrast Data Centre (HC-DC), supported by INSU/CNRS and ESO. This research has made use of the VizieR catalogue access tool, CDS, Strasbourg, France \citep{Ochsenbein2000}. The original description of the VizieR service was published in \citet{Ochsenbein2000}. This research had made use of the SIMBAD database (operated at CDS, Strasbourg, France; \citep{Wenger2000}). P.G., B.G., and N.B.O. acknowledges support in part from NASA EPSCoR 80NNSC22M0063, NSF AST 2319415, and NASA EPSCoR 80NSSC24M0112.

% We thank Johan Olofsson for his assessment of the SPHERE data quality and for discussions regarding saturation effects and observational limitations. We thank Jason Aufdenberg and Tomomi Otani for feedback on the manuscript. 

% %____________Reference Section_________________________________

% \FloatBarrier
% \clearpage

% \bibliographystyle{aasjournal}
% \bibliography{ref}
% \end{document}

%%%%%%%%%%%%%%%%%%%%%%%%%%%%%%%%
% \newpage
%_____________Appendix_________________________________________

\appendix
\vspace{-0.3cm}
\section{Best-Fit Temperature for the SED}
This appendix summarizes the broadband SED fitting results for all stars in the sample. We list the effective temperature from the literature, the best-fit blackbody temperature obtained from $\chi^2$ minimization, and the corresponding $\chi^2$ values. The fitted temperatures represent effective broadband color temperatures derived from heterogeneous catalog photometry.

\begin{table}[h]
\centering
\caption{Blackbody temperatures derived from SED fits in the optical--near-IR regime (0.39--2.2~$\mu$m). The fitted temperature represents an effective broadband color temperature and may differ from literature $T_{\rm eff}$ due to the absence of explicit extinction correction and the use of heterogeneous catalog photometry. Uncertainties correspond to the 1$\sigma$ confidence interval from the likelihood profile.}
\label{tab:sed_temps}

\begin{tabular}{lccc}
\hline
Star & $T_{\rm eff,ref}$ (K) & $T_{\rm BB}$ (K) & $\chi^2$ \\
\hline
HD38915  & 8100 & $7250^{+60}_{-50}$ & 129.85\\
HD50877  & 3900 & $3550^{+30}_{-30} $& 85.42\\
HD52877  & 3750 & $3280^{+20}_{-20} $& 168.71\\
HD58526  & 5380 & $5280^{+90}_{-80} $& 560.03\\
HD67594  & 5210 & $4730^{+20}_{-30} $& 16.88\\
HD74395  & 5370 & $5140^{+60}_{-50} $& 50.90\\
HD209750 & 5210 & $4710^{+30}_{-40} $& 365.21\\
\hline
\end{tabular}
\end{table}

%____________Reference Section_________________________________

\FloatBarrier
% \clearpage

% \bibliography{ref}
% \bibliographystyle{aasjournal}
\bibliographystyle{aasjournal}
\bibliography{ref}
\end{document}